\documentclass[a4paper,flenq,usenatbib]{mnras}

\usepackage{newtxtext,newtxmath}

\usepackage[T1]{fontenc}
\usepackage{ae,aecompl}

\usepackage{graphicx}	
\usepackage{amsmath}	
\usepackage{amssymb}	
\usepackage{siunitx}	

\usepackage[usenames,dvipsnames]{color}
\definecolor{purple}{rgb}{0.5,0,0.5}

\newcommand{\lrf}[2]{\left(\frac{#1}{#2}\right)}			
\newcommand{\lrfe}[3]{\left(\frac{#1}{#2}\right)^{#3}}		
\newcommand{\eqr}[1]{equation\ \eqref{#1}}					

\numberwithin{equation}{section}
\title[Tidal heating, stellar irradiation of hot Jupiters]{Tidal heating and stellar irradiation of Hot Jupiters}

\author[A. S. Jermyn]{Adam S. Jermyn,$^{1}$\thanks{E-mail: adamjermyn@gmail.com}
Christopher A. Tout,$^{1}$
Gordon I. Ogilvie,$^{2}$
\\
$^{1}$Institute of Astronomy, University of Cambridge, Madingley Rd, Cambridge CB3 0HA, UK\\
$^{2}$Department of Applied Mathematics and Theoretical Physics, University of Cambridge, Madingley Rd, Cambridge CB3 0WA, UK\\
}

\date{Accepted 2017 March 31. Received 2017 March 30; in original form 2016 November 9}

\pubyear{2017}

\begin{document}
\label{firstpage}
\pagerange{\pageref{firstpage}--\pageref{lastpage}}
\maketitle

\begin{abstract}
We study the interaction between stellar irradiation and tidal heating in gaseous planets with short orbital periods.
The intentionally simplified atmospheric model we employ makes the problem analytically tractable and permits the derivation of useful scaling relations.
We show that many tidal models provide thermal feedback, producing interior radiative zones and leading to enhanced g-mode dissipation with a wide spectrum of resonances.
These resonances are dynamically tuned by the thermal feedback, and so represent a novel form of thermomechanical feedback, coupling vibrational modes to the very slow thermal evolution of the planet.
We then show that stellar irradiation allows the heat produced by these modes to be trapped at depth with high efficiency, leading to entropy increase in the central convective region, as well as expansion of the planet's radius sufficient to match observed swelling.
We find that thermally driven winds play an essential role in this process by making the thermal structure of the atmosphere spherically symmetric within a few scale heights of the photosphere.
We characterise the relationship between the swelling factor, the orbital period and the host star and determine the timescale for swelling.
We show that these g-modes suffice to produce bloating on the order of the radius of the planet over $\mathrm{Gyr}$ timescales when combined with significant insolation and we provide analytic relations for the relative magnitudes of tidal heating and insolation.
\end{abstract}




\section{Introduction}

\begin{figure}
\centering
\includegraphics[width=0.5\textwidth]{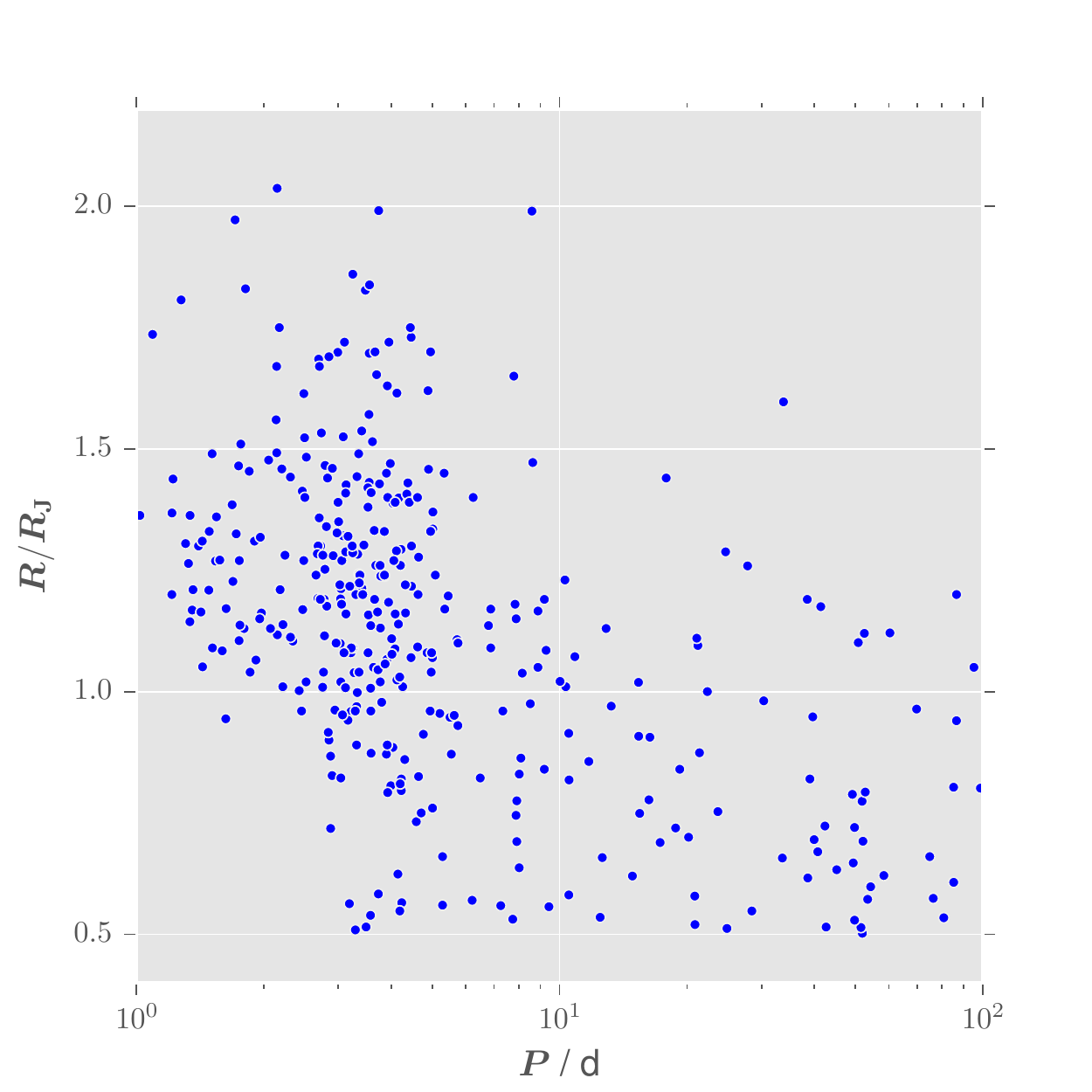}
\caption{Known population of short-period planets with radii near that of Jupiter.}
\label{fig:population}
\end{figure}

In recent data from Kepler and ground-based followups, there is evidence for a large population of hot Jupiters which are substantially inflated relative to their degenerate radii\ \citep{2012MNRAS.426..739H,2013ApJ...768...14W,2012AJ....144..139H}.
The radii and periods of known members of this population as well as of the broader Jupiter-sized population are shown in Fig.~\ref{fig:population}\ \citep{1211.7121}.
There is an apparent split in the observed population around periods of $\SI{10}{d}$, such that planets with longer periods are generally not inflated while those with shorter periods are often substantially inflated.
Importantly, planets at or above $2 R_\mathrm{J}$ must be inflated relative to their degenerate radii, otherwise their implied masses would make them stars\ \citep{1991ARA&A..29..163S}.
Importantly, Jupiter has approximately the maximum radius for an unheated gas giant, so planets at $2 R_\mathrm{J}$ must be bloated regardless of their mass.
In order to achieve this level of expansion, the central convection zone must be heated considerably relative to what would be expected as a result of the residual heat of formation\ \citep{2016ApJ...818....4L}, and there is evidence of planets re-inflating after cooling down\ \citep{1609.02767}.
Complicating this is the thermodynamic requirement that heat flows only from hot to cold, not in the reverse fashion, and absent an internal heat source there is no means for blocked heat transport to heat the interior.
This, combined with the expectation that temperature increases towards the core of the planet, means that any change in temperature at depth must be due to heat generated at or deeper than the point of interest.

A variety of mechanisms have been suggested to generate deep heating, with ohmic processes\ \citep*{2011ApJ...738....1B,2013ApJ...772...76S} and tidal dissipation\ \citep{2013arXiv1304.4121S,2009ApJ...702.1413M} among the more popular models.
Given that heat cannot flow from the surface of the planet into its core, the stellar flux is often neglected.
However, somewhat surprisingly, the observed radii correlate strongly with the incident stellar flux, so that this flux may play a role in the inflation process\ \citep{2016ApJ...818....4L}.
Confounding this analysis is the fact that stellar flux is not independent of orbital period.
So a theory of hot Jupiter inflation must separately handle the effects of orbital period and incident flux, particularly when dealing with tidal heating.

We investigate the effects of stellar flux on the structure of an internally heated hot Jupiter, making few assumptions about the nature or profile of the heating and considering the effects of wind redistribution.
We show that the stellar flux acts to modulate the rate at which heat escapes from the planet.
We then investigate the feedback that this heating produces on the thermal structure of the planet and show that a wide variety of realistic heating profiles gives rise to interior radiative zones.
These zones migrate within the planet on thermal timescales, giving a broad and dynamically tuned spectrum of g-mode resonances which dissipate heat tidally in the planet.
We then show that these modes suffice to produce the observed bloating.
Finally we predict the relation between stellar flux, orbital period and planetary radius.

\begin{figure}
\centering
\includegraphics[width=0.45\textwidth]{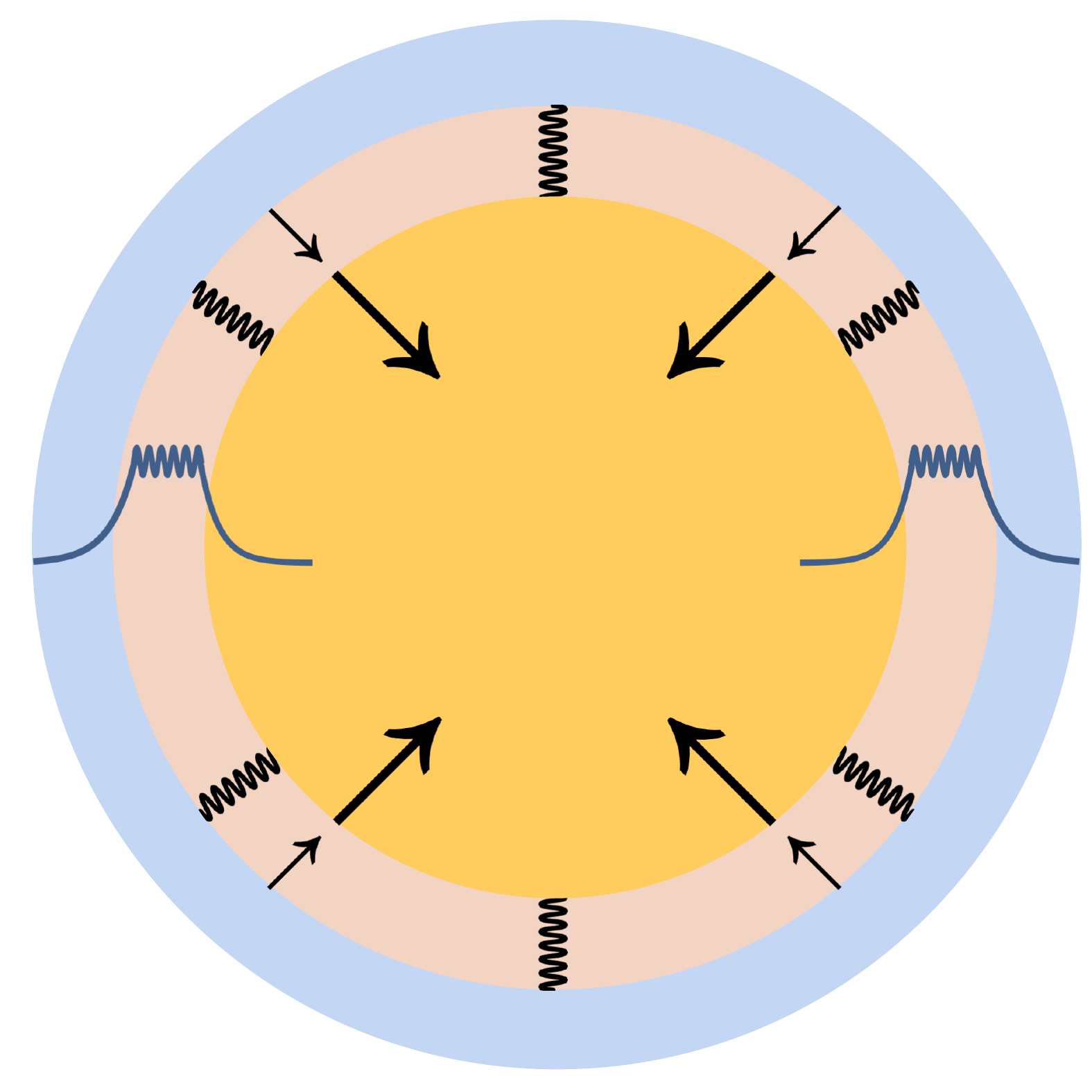}
\caption{Schematic of the proposed thermomechanical feedback mechanism. The upper convective layer (blue), radiative layer (beige), and inner convective layer (yellow) are shown as concentric shells. The boundaries of the radiative layer are moving inward at different rates, allowing the zone to resize. Profiles of g-modes (dark blue) are shown along the equator and schematically depicted at other latitudes.}
\label{fig:schematic}
\end{figure}

The new thermomechanical feedback mechanism we propose, shown schematically in Fig.~\ref{fig:schematic}, underscores the importance of considering planets as dynamical objects with complex behaviours coupling wildly different timescales.
Vibrational effects with periods ranging from seconds to days can have a tremendous impact on thermal evolution over millions of years, and that thermal evolution in turn feeds back into the vibrational modes, creating a dynamically tuned spectrum which can ultimately determine the large-scale structure of the planet.

\section{Isotropic Planetary Structure}

\begin{figure}
\centering
\includegraphics[width=0.5\textwidth]{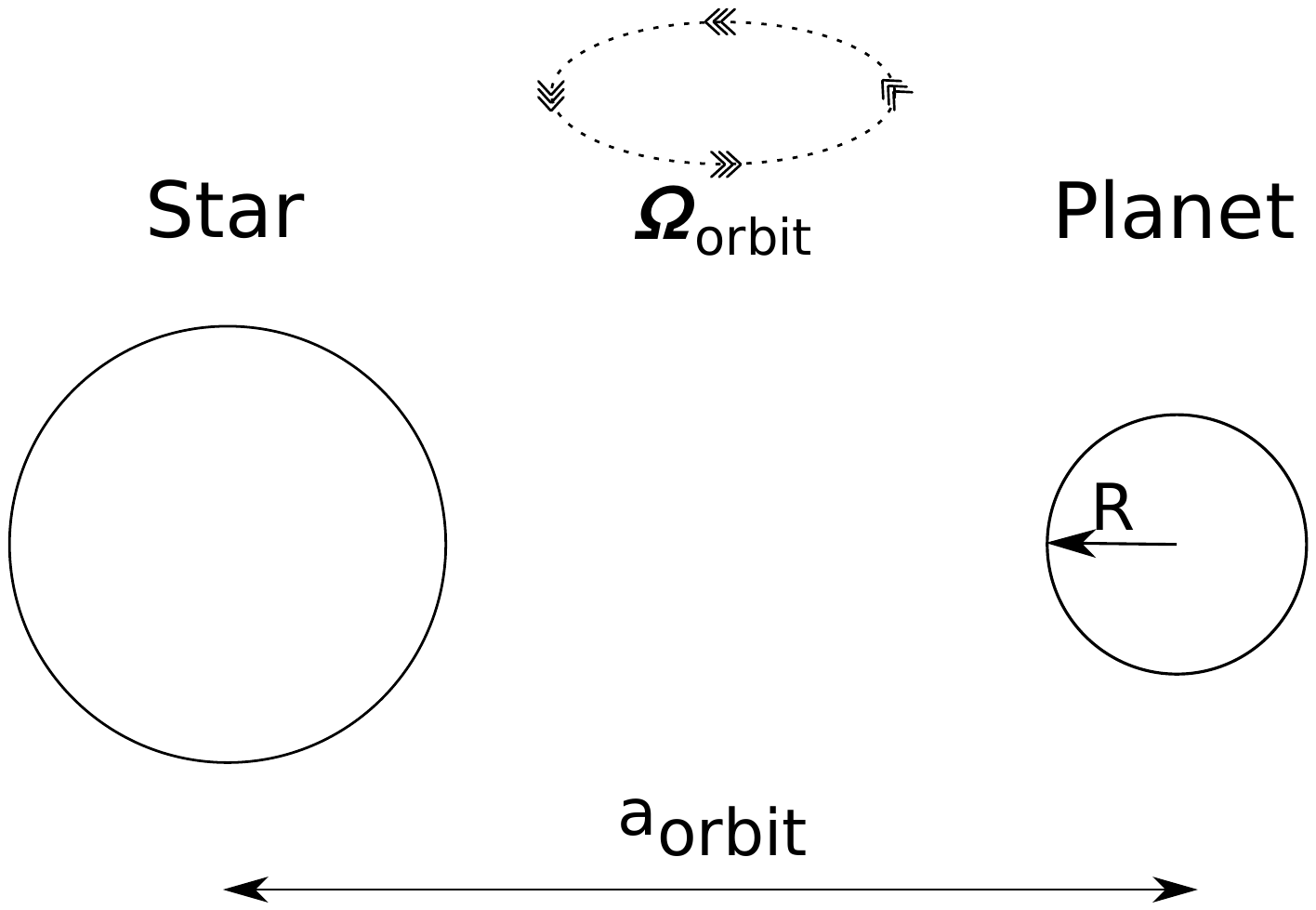}
\caption{Orbital configuration of the planet and its host star.}
\label{fig:orbit}
\end{figure}

\begin{figure}
\centering
\includegraphics[width=0.45\textwidth]{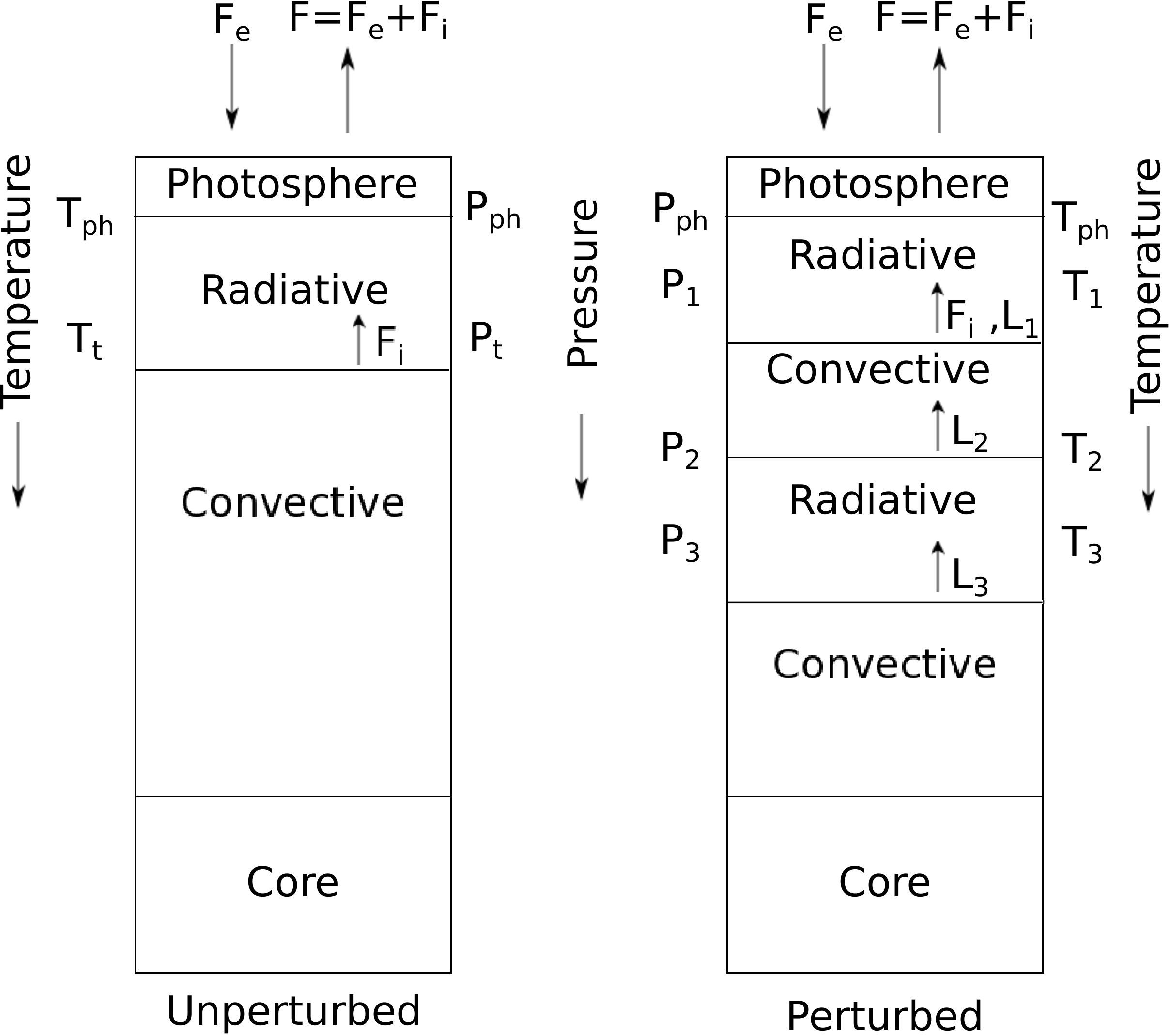}
\caption{Thermodynamic structure of the planet. The unperturbed structure is shown on the left, while the heated (perturbed) structure is shown on the right.}
\label{fig:thermo}
\end{figure}

We discuss the structure of a planet isotropically illuminated by flux $F_\mathrm{e}$ from its host star.
Fig.~\ref{fig:orbit} shows the orbital configuration of the planet-star system and Fig.~\ref{fig:thermo} shows the thermodynamic structure of the planet with the relevant variables defined schematically.
For the deep interior of the planet we adopt the analytic brown dwarf structure of\ \citet{1991ARA&A..29..163S}:
\begin{align}
\label{eq:structure1}
\psi &\equiv \frac{k_\mathrm{B} T}{E_\mathrm{F}} = \num{8e-6}\mu_{\mathrm{e}}^{2/3}\lrfe{\rho}{\si{g.cm^{-3}}}{-2/3}\lrf{T}{\si{K}},\\
\label{eq:structure2}
R_0 &= \SI{2.8e9}{cm}\lrfe{M}{M_{\sun}}{-1/3}\mu_{\mathrm{e}}^{-5/3},\\
\label{eq:structure3}
r &= R_0\left(1+\psi+\frac{\psi^2}{1+\psi}\right),\\
\label{eq:structure4}
P &= \SI{e13}{erg.cm^{-3}}\mu_{\mathrm{e}}^{-5/3}\lrfe{\rho}{\si{g.cm^{-3}}}{5/3}\lrf{r}{R_0},\\
\intertext{and}
\label{eq:structure5}
\nabla_\mathrm{a} &\equiv \left.\frac{\partial \ln T}{\partial \ln P}\right|_s = \frac{2}{5},
\end{align}
where $\psi$ is the electron degeneracy parameter, $\rho$ is the density, $T$ is the temperature, $R_0$ is the degenerate radius, $r$ is the radial coordinate, $P$ is the pressure, $k_\mathrm{B}$ is the Boltzmann constant, $E_\mathrm{F}$ is the Fermi energy, $M$ is the mass of the planet and $\nabla_\mathrm{a}$ is the adiabatic temperature gradient.
These relations effectively parametrise a $\gamma=5/3$ adiabatic atmosphere, accounting for electron degeneracy at high pressures.
We assume solar composition in this paper, so that the mean molecular weight of electrons $\mu_\mathrm{e} \approx 1.15$.
In addition, we take $R$ to be the radius of the planet.
For convenience, we define the parameters
\begin{align}
\label{eq:MATHR}
\mathcal{R} &\equiv R/R_0,\\
\intertext{and}
\label{eq:MATHM}
\mathcal{M} &\equiv M/M_{\mathrm{J}},
\end{align}
where $M_{\mathrm{J}}=1.838\times 10^{30}\mathrm{g}$ is the mass of Jupiter.
We expect the gas line opacity to dominate in hot atmospheres, so we use this as fiducial and define\ \citep{1991ARA&A..29..163S}
\begin{equation}
	\kappa_0 \equiv \SI{e-2}{cm^2.g^{-1}	}.
\end{equation}

We connect the top of the convection zone to the photosphere with a radiative zone at transition pressure $P_\mathrm{t}$.
The photospheric temperature is given by
\begin{equation}
T_\mathrm{ph} = \lrfe{F}{\sigma}{1/4},
\end{equation}
where $F$ is the total flux leaving the planet's atmosphere and $\sigma$ is the Stefan-Boltzmann constant.
This flux may be divided into two components, as
\begin{equation}
F = F_\mathrm{i} + F_\mathrm{e},
\end{equation}
where $F_\mathrm{i}$ is the heat arriving from the planet's interior and $F_\mathrm{e}$ is the heat arriving from the host star.
When $F_\mathrm{e}$ is large relative to the flux which would escape through the planet's natural cooling the photospheric temperature is determined entirely by $F_\mathrm{e}$, such that
\begin{equation}
T_\mathrm{ph} = \lrfe{F_\mathrm{e}}{\sigma}{1/4}.
\label{eq:largeanistemp}
\end{equation}
The escaping flux $F_\mathrm{i}$ from the planet's core is just the flux which escapes from the convection zone.
This may be generated by gravitational contraction, radioactive decay or by a decrease in the interior entropy; here we generally assume the last of these to be the dominant source of interior flux.

The radiative gradient is given by
\begin{equation}
\nabla_\mathrm{r} = \frac{3\kappa P L}{64\pi GM \sigma T^4},
\label{eq:gradR}
\end{equation}
where $L$ is the luminosity driving the gradient.
At the convective-radiative boundary we have
\begin{equation}
\nabla_\mathrm{a} = \nabla_\mathrm{r} = \frac{3\kappa P_\mathrm{t} L_\mathrm{i}}{64\pi GM \sigma T_\mathrm{t}^4},
\label{eq:radad}
\end{equation}
where 
\begin{equation}
L_\mathrm{i} = 4\pi R^2 F_\mathrm{i}.
\end{equation}
If we take $\kappa$ to be a power law in both $P$ and $T$ of the form
\begin{equation}
\kappa = \kappa_0 T^a P^b
\label{eq:kappaPWR}
\end{equation}
then
\begin{equation}
\nabla_\mathrm{a} = \frac{3\kappa_0 P_\mathrm{t}^{1+b} L_\mathrm{i}}{64\pi GM \sigma T_\mathrm{t}^{4-a}}.
\end{equation}
Above the transition we have
\begin{equation}
\nabla_\mathrm{r} = \nabla_\mathrm{a}\lrfe{P}{P_\mathrm{t}}{1+b}\lrfe{T}{T_\mathrm{t}}{a-4} = \frac{d\ln T}{d\ln P}.
\end{equation}
Integrating from the photosphere to the transition yields
\begin{equation}
1 - \lrfe{T_\mathrm{ph}}{T_\mathrm{t}}{4-a} = \frac{4-a}{1+b} \nabla_\mathrm{a}\left(1 - \lrfe{P_\mathrm{ph}}{P_\mathrm{t}}{1+b}\right).
\end{equation}
The photosphere pressure is generally much lower than the transition pressure so
\begin{equation}
\frac{T_\mathrm{ph}}{T_\mathrm{t}} = \left(1 - \frac{4-a}{1+b} \nabla_\mathrm{a}\right)^{\frac{1}{4-a}},
\end{equation}
which has a solution if and only if
\begin{equation}
1 - \frac{4-a}{1+b} \nabla_\mathrm{a} \geq 0.
\label{eq:kappaTPcond}
\end{equation}
If $b > -1$ and $a > 4$ or $b<-1$ and $a < 4$ or $b > \frac{3-2a}{5}$ and $a < 4$ this is satisfied.
There are other conditions under which it is satisfied, but these are the most relevant common cases.
The exponents are generally of order unity and the result is raised to a small power so $T_\mathrm{ph} \approx T_\mathrm{t}$.
This agrees with other analyses, which have found that in Jupiter-like planets, $T_\mathrm{ph} < T_\mathrm{t} < 1.5 T_\mathrm{ph}$\ \citep{0004-637X-803-2-111}.
The precise temperature ratio depends on the nature of the opacity function, so we simply take $T_\mathrm{t} = 2^{1/4} T_\mathrm{ph}$ (the Eddington closed grey body) as representative.
If this holds and the flux from the star is dominant then
\begin{equation}
F_\mathrm{i} =  \frac{16\sigma g \nabla_\mathrm{a} T_{\mathrm{t}}^4}{3\kappa P_\mathrm{t}} = \lrf{32 g \nabla_\mathrm{a}}{3 \kappa P_\mathrm{t}}F_\mathrm{e}.
\label{eq:fluxrat0}
\end{equation}
Eliminating $\mu_e$ between equations\ \ref{eq:structure1} and \ \ref{eq:structure4}, we may write $P_\mathrm{t}$ in terms of $T_\mathrm{t}$ at $r\approx R$ and $R_0$, such that
\begin{equation}
P_\mathrm{t} = \SI{e13}{erg.cm^{-3}} \lrf{R}{R_0}\lrfe{T_\mathrm{t}}{\si{K}}{5/2}\lrfe{\psi}{\num{8e-6}}{-5/2}.
\label{eq:pt1}
\end{equation}
From \eqr{eq:structure3} and the definition of $R_0$ (\eqr{eq:MATHR}) we find
\begin{align}
\psi &= \frac{1}{4}\left(\frac{R}{R_0}-2+ \sqrt{\lrfe{R}{R_0}{2}+4\lrf{R}{R_0} - 4}\right)\\
&=\frac{1}{4}\left(\mathcal{R}-2+ \sqrt{\mathcal{R}^2+4\mathcal{R} - 4}\right),
\end{align}
where we have taken the positive root because $\psi>0$ and $R > R_0$.
Note that this is always of order unity, so to good approximation the majority of the variation in $P_\mathrm{t}$ comes from the $R$ and $T_\mathrm{t}$ dependence in \eqr{eq:pt1}.

If the convective-radiative transition occurs at a shallow point in the atmosphere the corresponding column density is just
\begin{equation}
\Sigma_\mathrm{t} \approx \frac{P_\mathrm{t}}{g} = \SI{4e9}{g.cm^{-2}}\mathcal{R}^{3} \mathcal{M}^{-5/3}\lrfe{T_\mathrm{t}}{\si{K}}{5/2}\lrfe{\psi}{\num{8e-4}}{-5/2},
\end{equation}
again with $\mathcal{M} = M/M_\mathrm{J}$.
Eliminating $T_\mathrm{t}$ in favour of $F_\mathrm{e}$ and using $T_{\sun} = 5777\mathrm{K}$ we find
\begin{align}
\Sigma_\mathrm{t} \approx \SI{2e6}{g.cm^{-2}}\mathcal{R}^{3}\mathcal{M}^{-5/3}\lrfe{F_\mathrm{e}}{F_{\sun}}{5/8}\psi^{-5/2}.
\end{align}
Inserting this result into \eqr{eq:fluxrat0} we obtain
\begin{align}
\frac{F_{\mathrm{i}}}{F_{\mathrm{e}}} &= \frac{32g\nabla_{\mathrm{a}}}{3\kappa P_{\mathrm{t}}}= \num{2e-4}\mathcal{M}^{5/3}\lrfe{F_{\mathrm{e}}}{F_{\sun}}{-5/8}\lrfe{\kappa}{\kappa_0}{-1}\frac{\psi^{5/2}}{\mathcal{R}^3}.
\label{eq:fluxrat1}
\end{align}

As one final manipulation, we wish to put our equations in terms of the stellar luminosity and orbital radius.
The stellar luminosity is related to the external flux $F_{\mathrm{e}}$ by
\begin{equation}
4 \pi R^2 F_{\mathrm{e}} = \frac{\pi R^2 L_\star}{4\pi a_\mathrm{orbit}^2},
\label{eq:orbitalFlux}
\end{equation}
where $L$ is the stellar luminosity, $a_\mathrm{orbit}$ is the orbital radius of the planet.
The factor of $\pi R^2$ on the right-hand side is just the cross-section of the planet as seen from the star, while the factor of $4\pi R^2$ on the left-hand side reflects the definition of $F_\mathrm{e}$ as an average over the surface of the planet.
So
\begin{equation}
F_{\mathrm{e}} = \frac{L}{16\pi a_\mathrm{orbit}^2}.
\end{equation}
Comparing with the Sun we find
\begin{equation}
\frac{F_{\mathrm{e}}}{F_{\sun}} = \frac{1}{4}\lrf{L_\star}{L_{\sun}}\lrfe{a_\mathrm{orbit}}{R_{\sun}}{-2}.
\label{eq:fefsun}
\end{equation}
Thus
\begin{align}
\frac{F_{\mathrm{i}}}{F_{\mathrm{e}}} &= \num{5e-4}\mathcal{M}^{5/3}\lrfe{L_\star}{L_{\sun}}{-5/8}\lrfe{a_\mathrm{orbit}}{R_{\sun}}{5/4}\lrfe{\kappa}{\kappa_0}{-1}\frac{\psi^{5/2}}{\mathcal{R}^3}.
\label{eq:fluxrat3}
\end{align}
Importantly, the exponent on the luminosity is greater than $-1$.
This means that while the ratio of escaping to incident flux decreases with increasing stellar flux, the total escaping flux increases.
This conclusion is dependent primarily on how strongly the ratio $T_\mathrm{t}/T_\mathrm{ph}$ varies with $F_\mathrm{e}$, which in turn depends on the form of the opacity function.
In particular, it does not generally hold at extremely high temperatures where the gas line opacity ceases to dominate and Kramers-like rules take over.
For brown dwarfs and hot Jupiters, however, this variation is small and should not pose a problem.
It is also useful to compute the transition column density
\begin{equation}
\Sigma_\mathrm{t} \approx \frac{P_\mathrm{t}}{g} = \SI{8e5}{g.cm^{-2}}\mathcal{R}^{3}\mathcal{M}^{-5/3}\lrfe{L_\star}{L_{\sun}}{5/8}\lrfe{a_\mathrm{orbit}}{R_{\sun}}{-5/4}\psi^{-5/2}.
\label{eq:sigt}
\end{equation}
This is small enough that the shallow approximation is not bad.

\section{Angular Temperature Distribution}
The planets under consideration are generally highly insolated.
This can lead to significant temperature differences between the day and night sides, particularly if the planet is tidally locked.
In this section we show that winds suffice to make the thermal structure of the atmosphere spherically symmetric at depth even when there is a large temperature difference at the photosphere.
This allows us to treat the structure of the planet as spherically symmetric where tidal effects are most prominent.

Consider a wind driven from one side of the planet to the other along isobars with characteristic velocity $\boldsymbol{\varv}$.
Suppose further that the character of this wind changes in the vertical direction over distances of order the pressure scale height $h$ and that it changes in the horizontal direction over distances of order the planet's radius.
The specific force due to shear in the vertical direction is
\begin{equation}
\boldsymbol{F}_\mathrm{v} = \nu_\mathrm{v} \frac{\partial \boldsymbol{\varv}}{\partial r},
\end{equation}
where $\nu_\mathrm{v}$ is the viscosity for a circumferential flow shearing in the vertical direction.
The corresponding power dissipated is
\begin{equation}
\mathcal{P}_\mathrm{v} = \frac{\partial \boldsymbol{\varv}}{\partial r}\cdot\boldsymbol{F}_\mathrm{v} = \nu_\mathrm{v} \lrfe{\partial \varv}{\partial r}{2}.
\end{equation}
Likewise, the force due to shear in the horizontal direction is
\begin{equation}
\boldsymbol{F}_\mathrm{h} = \nu_\mathrm{h} \frac{\partial \boldsymbol{\varv}}{\partial \xi}, 
\end{equation}
where $\nu_\mathrm{h}$ is the viscosity for a circumferential flow shearing in the other circumferential direction and $\xi$ is a coordinate along the flow.
The corresponding power dissipated is
\begin{equation}
\mathcal{P}_\mathrm{h} = \frac{\partial \boldsymbol{\varv}}{\partial \xi}\cdot\boldsymbol{F}_\mathrm{h} = \nu_\mathrm{v} \lrfe{\partial \varv}{\partial \xi}{2}.
\end{equation}
The total power dissipated is then
\begin{equation}
\mathcal{P} = \mathcal{P}_\mathrm{v} + \mathcal{P}_\mathrm{h} \approx \varv^2 \left[\frac{\nu_\mathrm{v}}{h^2} + \frac{\nu_\mathrm{h}}{r^2}\right],
\label{eq:winddissip}
\end{equation}
where we have approximated the velocity derivatives with the velocity magnitude and the relevant scale heights, the pressure scale height $h$ in the vertical direction and the radius $r$ in the horizontal.
We have also simplified the viscosity from a rank-4 tensor to two scalars, so this relation ought only to be interpreted as an order of magnitude of the power.

To determine $v$, we now match this power to the work which the wind may extract as a heat engine.
We are interested in cases where the temperature difference between the two sides is large so the efficiency of the heat engine is of order unity even if diffusive losses make it irreversible.
We may neglect diffusive losses because we have taken the microscopic thermal diffusivity to be small on the relevant scales.
So we may write the specific rate of work as
\begin{equation}
\mathcal{W} = c_\mathrm{p} \boldsymbol{\varv} \cdot\nabla T \approx c_\mathrm{p} \varv \frac{\Delta T}{\pi r},
\end{equation}
where
\begin{equation}
c_\mathrm{p} = 5 R_\mathrm{gas}/2,	
\end{equation}
for a monatomic ideal gas, is the specific heat at constant pressure.
This is simply the specific heat which is transported from one side of the planet to the other.
In our case
\begin{align}
\Delta T &\equiv T_\mathrm{day}-T_\mathrm{night}\\
\intertext{and}
T &\equiv \frac{1}{2}\left(T_\mathrm{day}+T_\mathrm{night}\right),
\end{align}
so $T$ refers to the average temperature while $\Delta T$ refers to the temperature difference.
By definition, $\Delta T/T \leq 2$.
In the most extreme case this gives
\begin{equation}
\mathcal{W} \approx  c_\mathrm{p} \varv \frac{\Delta T}{\pi r} \approx \frac{c_\mathrm{p} T v}{\pi r}\lrf{\Delta T}{T}\approx \frac{5c_\mathrm{s}^2 v}{2 \gamma \pi r},
\end{equation}
where
\begin{equation}
c_{\mathrm{s}} = \sqrt{\frac{\gamma R_\mathrm{gas} T}{\mu}}	
\end{equation}
is the adiabatic sound speed and $\mu$ is the mean molecular weight.
Equating the rate of work and power gives
\begin{equation}
c_{\mathrm{s}}^2  = \frac{2\gamma}{5}\pi r \varv\left[\frac{\nu_\mathrm{v}}{h^2} + \frac{\nu_\mathrm{h}}{r^2}\right].
\label{eq:workbalance}
\end{equation}
To proceed further we must examine the forms of $\nu_\mathrm{v}$ and $\nu_\mathrm{h}$.
The nature of the viscosity differs between stably stratified and buoyantly unstable zones, so we must determine which of these are relevant and treat them separately.

We begin with radiative zones.
In a stably stratified region the two viscosities differ because of Richardson stabilisation, an effect which limits the scale of turbulence in the vertical direction by means of a buoyant restoring force\ \citep*{2007AtScL...8...65G}.
A straightforward prescription for the viscosities in this context is
\begin{align}
\label{eq:nuh}
\nu_{\mathrm{h}} &\approx \varv r \\
\nu_{\mathrm{v}} &\approx \varv^2\left(\frac{\alpha + \nu_{\mathrm{h}}}{g h(\nabla_\mathrm{a}-\nabla)}\right),
\end{align}
where $\alpha$ is the microscopic thermal diffusivity\ \citep{2004A&A...425..243M}.
Generally we expect $\alpha$ to be small compared to $\nu_{\mathrm{h}}$ because horizontal radiative transfer is inefficient., so we may neglect $\alpha$ and write
\begin{equation}
\nu_{\mathrm{v}} = \varv^2\left(\frac{\nu_{\mathrm{h}}}{g h(\nabla_\mathrm{a}-\nabla)}\right) = \frac{\varv^3 r}{g h (\nabla_\mathrm{a} - \nabla)}.
\end{equation}
By the Schwarzschild criterion $\nabla < \nabla_\mathrm{a}$ in a stably stratified zone.
In general we expect radiative transport to be efficient far from the zone boundaries, so we take $\nabla \ll \nabla_{\mathrm{a}}$ in most of such a zone.
Using this we write
\begin{equation}
\nu_{\mathrm{v}} = \frac{v^3 r}{g h \nabla_\mathrm{a}}.
\end{equation}
Now making use of
\begin{equation}
g h = g \left|\frac{dr}{d\ln P}\right| = gP\left|\frac{dr}{dP}\right| = \frac{gP}{g\rho} = \frac{P}{\rho} = \gamma^{-1} c_\mathrm{s}^2
\label{eq:gl}
\end{equation}
we find
\begin{equation}
\nu_{\mathrm{\varv}} = \frac{\varv^3 r \gamma}{c_\mathrm{s}^2 \nabla_\mathrm{a}}.
\label{eq:nuv}
\end{equation}
Inserting equations\ \eqref{eq:nuh} and \ \eqref{eq:nuv} into \eqr{eq:workbalance} gives
\begin{equation}
c_\mathrm{s}^2  = \frac{2\gamma}{5}\pi r \varv\left[\frac{\varv^3 r \gamma}{c_\mathrm{s}^2 h^2 \nabla_\mathrm{a}} + \frac{\varv}{r}\right].
\end{equation}
This may be rearranged to
\begin{equation}
\frac{5}{2\pi\gamma} =  \frac{\gamma}{\nabla_{\mathrm{a}}}\lrfe{\varv}{c_\mathrm{s}}{4} \lrfe{r}{h}{2} + \lrfe{\varv}{c_\mathrm{s}}{2}.
\end{equation}
Solving gives
\begin{equation}
\lrfe{\varv}{c_\mathrm{s}}{2} = \frac{\nabla_{\mathrm{a}}h^2}{2\gamma r^2}\left[-1\pm \sqrt{1 + \frac{10 r^2}{\pi \nabla_{\mathrm{a}} h^2}}\right].
\end{equation}
The positive branch is the one of interest, because we have implicitly taken $\varv > 0$ in writing it as a magnitude.
In the upper regions of the planet's atmosphere $r \gg h$ so
\begin{equation}
\frac{\varv}{c_\mathrm{s}} \approx \frac{h}{r}\sqrt{\frac{5}{2\pi \gamma}}.
\end{equation}
Using \eqr{eq:gl} the rate at which heat is transported may be written as
\begin{equation}
\varepsilon = \mathcal{W} \approx \frac{5 c_{\mathrm{s}}^2 \varv}{2\pi\gamma r} \approx \frac{c_{\mathrm{s}}^3 h}{r^2}\left(\frac{5}{2\pi\gamma}\right)^{3/2} \approx \frac{\gamma c_{\mathrm{s}}^5}{g r^2}\left(\frac{5}{2\pi\gamma}\right)^{3/2}.
\end{equation}
The region of interest is shallow so $gr^2 \approx GM$ and
\begin{equation}
\varepsilon \approx \frac{\gamma c_{\mathrm{s}}^5}{GM}\left(\frac{5}{2\pi\gamma}\right)^{3/2}.
\end{equation}

The depth, as measured by column density $\Sigma_{\mathrm{i}}$, over which the winds make the flux distribution spherically symmetric is
\begin{equation}
\Sigma_\mathrm{i} = \frac{F_\mathrm{e}}{\varepsilon} \approx \frac{GM \sigma T_\mathrm{ph}^4}{\gamma c_s^5}\left(\frac{2\pi\gamma}{5}\right)^{3/2}.
\end{equation}
Evaluating the sound speed at the photosphere gives
\begin{equation}
\Sigma_\mathrm{i} \approx \SI{3e3}{g.cm^{-2}}\mathcal{M} \lrfe{T}{\SI{e3}{K}}{3/2},
\end{equation}
where $m_\mathrm{p}$ is the proton mass.
For comparison, the photosphere is at a depth of
\begin{equation}
\Sigma_\mathrm{ph} \approx \kappa^{-1} = \SI{e2}{g.cm^{-2}}\lrfe{\kappa}{\kappa_0}{-1}.
\end{equation}
Thus the temperature distribution becomes spherically symmetric deeper than the photosphere but shallower than the convective transition.
So we need not worry about the viscosity in convection zones. 

This remains valid as long as the planet rotates slowly relative to $v/R$, such that the characteristic scale of circumferential motion remains $R$ and is not reduced by Coriolis effects.
At short periods, where this condition is most in danger, the anisotropy is very large, such that $v \approx h c_{\mathrm{s}}/r$, and the surface temperature should be quite high because of insolation, such that $c_{\mathrm{s}} \approx \SI{3e5}{cm.s^{-1}}$.
In this regime, the period of a Jupiter-radius planet must be at least $\SI{30}{d}$ with $h/r \approx \num{e-2}$ or $\SI{3}{d}$ with $l/r\approx \num{e-1}$ for the Coriolis effect to be negligible.
Even at the shortest known periods of just under a day, the correction term is not too great and does not alter the conclusion that the temperature distribution becomes spherical above the convection zone, so we continue to use this approximation with the knowledge that it becomes worse as the period diminishes.

\section{Heated Thermal Structure}

In this section we work on timescales long compared to the adjustment of radiative or convective zones to thermal perturbations but short compared to the characteristic thermal timescale of the planet.
This is the instantaneous equilibrium approximation.
This separation of scales exists because the thermal timescale of the planet is set by the thermal content of the core, whereas the radiative and convection regions of interest are shallow zones with much less mass and at much lower temperatures.

The equations governing the luminosity of the planet as a function of mass coordinate are
\begin{align}
\label{eq:timedif}
\frac{\partial L}{\partial m} &= \varepsilon(m) - c_\mathrm{p} \frac{\partial S}{\partial t},\\
L(0) &= 0\\
\intertext{and}
L(M) &= L_\mathrm{i},
\end{align}
where $\varepsilon(m)$ is the specific energy generation by tides, radioactive decay and ohmic processes and the mass coordinate $m$ corresponds to the spherical shell containing mass $m$.
Note that mechanical expansion and contraction can generate energy, but in this coordinate system that generation provides no net contribution because it does not alter the specific entropy.

Thermodynamic consistency imposes the condition that heat travels from hot regions to cool ones.
Assuming that $T$ increases towards the core of the planet, this means that $L(m) \geq 0$ everywhere.
In a convective atmosphere, the thermal gradient is almost independent of the luminosity.
This follows because the luminosity is determined by the superadiabaticity of the thermal gradient, rather than by the gradient.
When convection is efficient, the convective zone is nearly isentropic so $L\propto(\nabla-\nabla_\mathrm{a})^{3/2}$\ \citep{kippenhahn1} and the atmosphere achieves significant scaling of luminosity with only small changes to $\nabla$.
As a result, the conditions on $L$ cannot generally be satisfied.
This means that radiative zones are generically needed as interfaces between convective regions.
More formally, we work in the limit of perfectly efficient convection, such that
\begin{equation}
T(P)P^{-\nabla_\mathrm{a}} = \mathrm{const.}
\label{eq:adiabat}
\end{equation}
We also make the assumption that the convective turnover time for any region of interest is much shorter than the time-scale over which thermal quantities change, such that convection may be assumed to enforce an instantaneous adiabatic law.

Now suppose that we perturb a planet by injecting luminosity $\Delta L$ somewhere below the radiative-convective boundary.
For $\Delta L \ll L_\mathrm{i}$, we may solve \eqr{eq:timedif} by simply reducing the luminosity escaping from deeper regions of the planet.
That is, $L_\mathrm{i}$ goes unchanged but the luminosity in regions deeper than the injection depth is reduced by $\Delta L$.
In the limit of very efficient convection (or large opacity), this adjustment holds until $\Delta L \approx L_\mathrm{i}$.
For $\Delta L > L_\mathrm{i}$ the adjustment still occurs, with the deep luminosity falling to the radiative luminosity at the adiabatic gradient, the minimum needed to maintain convection.
The difference is that in this case there is an excess of luminosity reaching the convective-radiative transition and this must be accounted for.
At the boundary we must have
\begin{equation}
\nabla_\mathrm{a} = \nabla_\mathrm{r} = \frac{3\kappa P_\mathrm{t} L_\mathrm{i}}{64\pi GM \sigma T_\mathrm{t}^4},
\label{eq:radad}
\end{equation}
which must remain satisfied when we perturb $L_\mathrm{i}$ so
\begin{equation}
\Delta \ln P_\mathrm{t} - 4\Delta \ln T_{\mathrm{t}} + \Delta \ln \kappa + \Delta \ln L_\mathrm{i} = 0.
\label{eq:dlnrel}
\end{equation}
If the transition temperature is similar to the photospheric temperature and if the radius does not change substantially owing to the perturbation $\Delta \ln L = \Delta \ln (L_\mathrm{i} + L_\mathrm{e}) = 4\Delta \ln T_{\mathrm{t}}$.
Because $F_\mathrm{e} \gg F_\mathrm{i}$ and $L_\mathrm{e}$ is fixed
\begin{equation}
4\Delta \ln T_{\mathrm{t}} \approx \frac{\Delta L_\mathrm{i}}{L_\mathrm{e}} \ll 1.
\end{equation}
So we may neglect the change in $T_\mathrm{t}$ and find
\begin{equation}
\Delta \ln P_\mathrm{t} + \Delta \ln \kappa + \Delta \ln L_\mathrm{i}= 0.
\label{eq:dpdl}
\end{equation}
We generally expect that, at fixed temperature, $\kappa$ rises as $P$ rises.
As a result, $P_\mathrm{t}$ must fall to satisfy this relation, so either the entropy of the central adiabat must rise or the adiabatic law must be broken somewhere in the planet.
The central entropy cannot rise unless either heat is being added at the core or the photosphere is hotter than the core, because heat cannot be forced to move up the temperature gradient.
Neither of these are generally the case so the adiabatic law must be broken.
As a result the planet must form an interior radiative zone.

\begin{figure}
\centering
\includegraphics[width=0.45\textwidth]{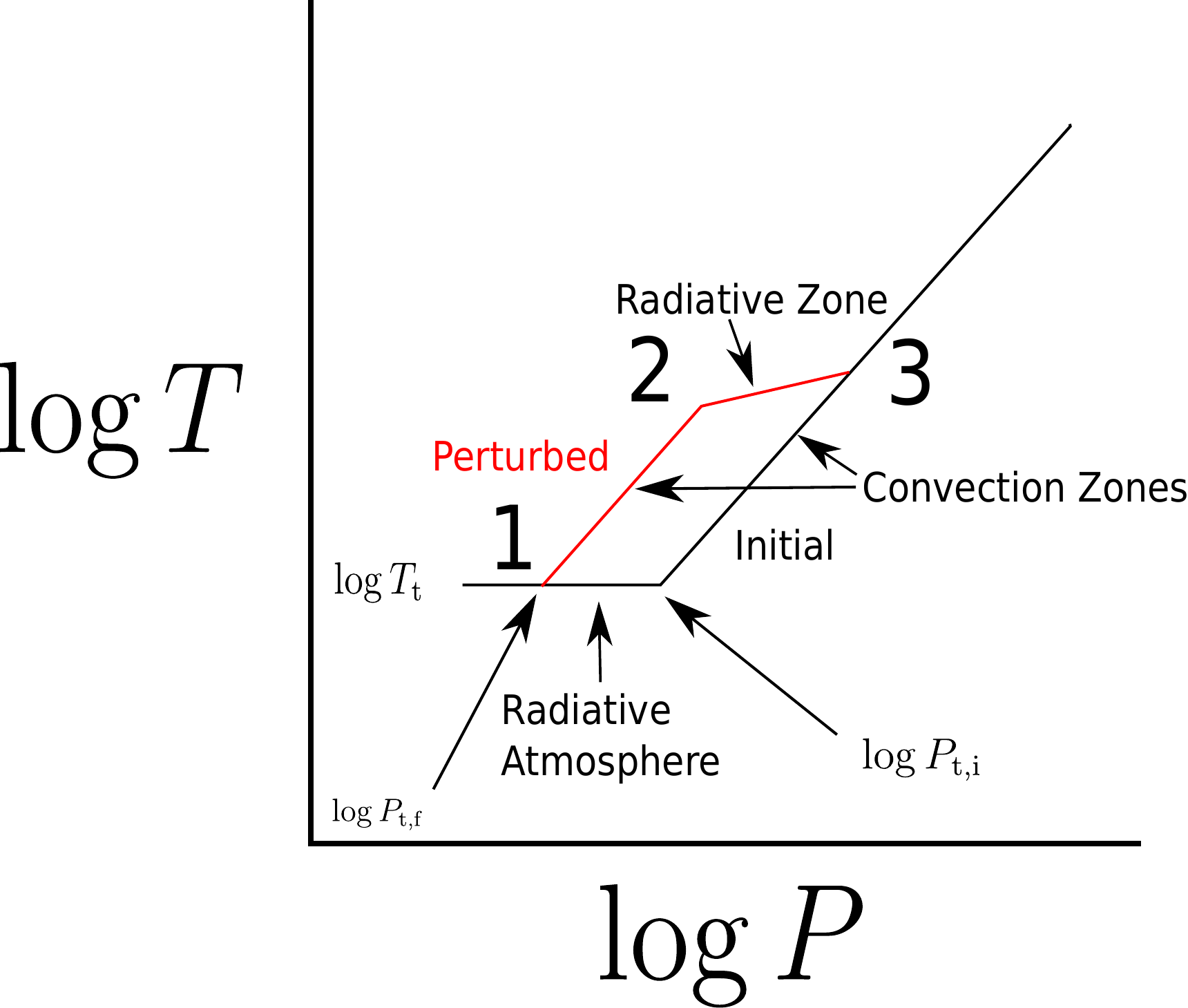}
\caption{Perturbed (red) and unperturbed (black) pressure-temperature profiles.}
\label{fig:pt}
\end{figure}

To characterise these radiative zones, let $P_1$ be the transition pressure between the surface radiative zone and the new convection zone, $P_2$ the transition pressure between this zone and the interior radiative zone and $P_3$ the transition pressure between this zone and the central adiabat.
The perturbed and unperturbed pressure-temperature structures are shown in Fig. \ref{fig:pt}.
Let $T_j$, $m_j$, $\kappa_j$ and $L_j$ be the corresponding temperature, mass coordinate, opacity and luminosity at each transition.
The new convection zone is adiabatic, so
\begin{equation}
\frac{T_1^{1/\nabla_\mathrm{a}}}{P_1} = \frac{T_2^{1/\nabla_\mathrm{a}}}{P_2}.
\end{equation}
Assuming that $m \approx M$, the condition \eqref{eq:radad} for transition between radiative and convective zones gives
\begin{equation}
P_1 \kappa_1 L_1 T_1^{-4} = P_2 \kappa_2 L_2 T_2^{-4} =  P_3 \kappa_3 L_3 T_3^{-4}.
\end{equation}
Finally, recalling that the central adiabat is fixed and that the new adiabat contains a point at the fixed temperature $T_\mathrm{t}$, we find
\begin{equation}
\frac{T_2^{1/\nabla_\mathrm{a}}}{P_2} = \frac{T_3^{1/\nabla_\mathrm{a}}}{P_3}\lrf{P_\mathrm{t,i}}{P_\mathrm{t,f}},
\label{eq:ent}
\end{equation}
where the subscripts $\mathrm{i}$ and $\mathrm{f}$ refer to the initial unperturbed and final perturbed system respectively.
Equation\ \eqref{eq:ent} thus expresses the entropy difference between the two adiabats.
Note that with these subscript definitions,
\begin{equation}
L_\mathrm{i,f} \equiv L_1.
\label{eq:l1def}
\end{equation}
Again with $\kappa \propto T^a P^b$ we may write equations\ \eqref{eq:dlnrel}~--~\eqref{eq:l1def} as a system of linear equations in the logarithms of temperature and pressure.
Solving this system yields
\begin{align}
\label{eq:trat1}
\ln \frac{T_1}{T_2} &= \frac{\nabla_\mathrm{a}}{w}\ln \frac{L_1}{L_2},\\
\label{eq:trat2}
\ln \frac{T_2}{T_3} &= \frac{\nabla_\mathrm{a}}{w}\ln \frac{L_{\mathrm{i,f}} L_2}{L_{\mathrm{i,i}} L_3},\\
\label{eq:prat1}
\ln \frac{P_1}{P_2} &= \frac{1}{w}\ln \frac{L_1}{L_2},
\intertext{and}
\label{eq:prat2}
\ln \frac{P_2}{P_3} &= \frac{1}{w}\ln \frac{L_{\mathrm{i,f}} L_2}{L_{\mathrm{i,i}} L_3} + \frac{1}{1+b}\ln\frac{L_{\mathrm{i,f}}}{L_{\mathrm{i,i}}},
\end{align}
where
\begin{equation}
	w \equiv (4-a)\nabla_\mathrm{a}-(1+b).	
\end{equation}
The transition temperature $T_1 \equiv T_\mathrm{t}$ is known from the unperturbed state so with equations \eqref{eq:trat1} and \eqref{eq:trat2} we may determine the remaining temperatures.
Likewise the unperturbed transition pressure $P_\mathrm{t,i}$ is known from the unperturbed state.
The perturbed transition pressure $P_1$ is related to the unperturbed by \eqr{eq:dpdl} so, with equations \eqref{eq:prat1} and \eqref{eq:prat2} we may determine the remaining pressures.

In equilibrium, the luminosities are related by
\begin{align}
L_1 &= L_2 + \int_{m_2}^{m_1} \epsilon(m) dm\\
\intertext{and}
L_2 &= L_3 + \int_{m_3}^{m_2} \epsilon(m) dm.
\end{align}
With these we can compute the luminosity ratios.
A consequence of \eqr{eq:prat1} is that the new convective zone is maintained by heat generation in between $P_1$ and $P_2$, or equivalently between $m_1$ and $m_2$, because this is what allows for $L_1 \neq L_2$.

The minimum luminosity required for convection may be calculated from \eqr{eq:radad} as
\begin{equation}
L_\mathrm{min} = L_\mathrm{i} \frac{\nabla_\mathrm{a}}{\nabla_\mathrm{r}}.
\end{equation}
Both sides of this equation are functions of pressure.
We generally expect that $\nabla_\mathrm{r}$ rises quickly towards the interior of the planet as convection becomes more efficient so $L_\mathrm{min}$ is a small fraction of $L_\mathrm{i}$.
This is actually guaranteed by \eqr{eq:kappaTPcond} so we expect that $L_\mathrm{min}$ is suppressed relative to $L_\mathrm{i}$ by a power-law in $P$ and may calculate
\begin{equation}
L_3 = \int_{0}^{m_3} \epsilon(m) dm + L_\mathrm{i} \frac{\nabla_\mathrm{a}}{\nabla_\mathrm{r}(P_\mathrm{inject})},
\label{eq:L3_1}
\end{equation}
where $P_\mathrm{inject}$ is the pressure inside which minimal luminosity is injected.

\section{Expansion}
\label{sec:expansion}

The expansion associated with changing the temperature profile of the planet is given by
\begin{equation}
\Delta V = \int_0^M \Delta\left(\rho^{-1}\right)dm.
\label{eq:dv0}
\end{equation}
In the limit where $\Delta R/R$ is small, $\Delta P/P$ is small at fixed $m$, so
\begin{equation}
	\Delta\left(\rho^{-1}\right) \approx \left.\frac{\partial T}{\partial \rho}\right|_{P}\Delta\left(T^{-1}\right) \approx \rho^{-1}\frac{\Delta T}{T} \approx \rho^{-1} \Delta \ln T.
\end{equation}
Substituting this into \eqr{eq:dv0} we find
\begin{equation}
	\Delta V \approx \int_0^M \rho^{-1}\Delta \ln T dm \approx \int_0^R 4\pi r^2 \Delta \ln T dr,
\end{equation}
where the coordinate $r$ refers to the unheated system.
The integration proceeds up to $R$ as an approximation, once more in the limit where $\Delta R/R$ is small.
When this is the case and when the majority of the heating occurs near the surface at $r \approx R$ this may be approximated by
\begin{equation}
\Delta R \approx \int_{R-\delta R}^{R} \Delta \ln T dr.
\label{eq:dr0}
\end{equation}
Now we may approximate $\Delta \ln T$ as $\ln (T_2/T_1)$.
The pressure depth over which this approximation (rather than $\Delta \ln T \approx 0$) is valid is $\Delta \ln P \approx \ln (P_1/P_\mathrm{t,i})$.
This corresponds to a physical depth of $h \ln (P_1/P_\mathrm{t,i})$, because $h$ is the characteristic scale of the thermal properties of the planet and hence sets the scale of the radiative zone which forms.
So we may approximate \eqr{eq:dr0} in terms of the heating parameters by
\begin{equation}
\Delta R \approx h \ln\frac{P_1}{P_{\mathrm{t,i}}} \ln\frac{T_2}{T_1}.
\label{eq:expansion0}
\end{equation}
With \eqr{eq:dpdl} we find
\begin{equation}
\Delta R = -\frac{h\nabla_\mathrm{a}}{\left(1+b\right)\left(w\right)} \ln \frac{L_{\mathrm{i,f}}}{L_\mathrm{i,i}}\ln \frac{L_1}{L_2}.
\end{equation}
For very deep zones, the relevant scale height is that near the base of the zone rather than the top, because the majority of the contribution to the integral comes from this region.
This may be taken into account by noting that the scale height at the base of the radiative zone is given by
\begin{equation}
h = -\frac{dr}{d\ln P} = \frac{k_B T_3}{\mu m_\mathrm{p} g}.
\label{eq:scaleheight}
\end{equation}
Inserting \eqr{eq:trat1} and \eqr{eq:trat2} we have
\begin{equation}
h = \frac{k_B T_1}{\mu m_\mathrm{p} g} \left(\frac{L_{\mathrm{i,f}}^2}{L_{\mathrm{i,i}} L_3} \right)^{-\frac{\nabla_\mathrm{a}}{w}}.
\end{equation}
Making use of $T_1 \approx T_\mathrm{ph}$, equations \eqref{eq:largeanistemp} and \eqref{eq:orbitalFlux} give us
\begin{align}
h &= \frac{k_B}{\mu m_\mathrm{p} g} \lrfe{L_\star}{4\pi \sigma a_\mathrm{orbit}^2}{1/4}\left(\frac{L_{\mathrm{i,f}}^2}{L_{\mathrm{i,i}} L_3} \right)^{-\frac{\nabla_\mathrm{a}}{w}}\\
\label{eq:scaleHeightExp}
&\approx 0.2R_\mathrm{J}\lrfe{L_\star}{L_{\sun}}{\frac{1}{4}}\lrfe{M_\star}{M_{\sun}}{-\frac{1}{6}}\lrfe{\tau_\mathrm{orbit}}{10\si{d}}{-\frac{4}{3}}\\
	&\times\lrfe{M}{M_\mathrm{J}}{-1}\lrfe{R}{R_\mathrm{J}}{2}\left(\frac{L_{\mathrm{i,f}}^2}{L_{\mathrm{i,i}} L_3} \right)^{-\frac{\nabla_\mathrm{a}}{w}},\nonumber
\end{align}
where $\tau$ is the orbital period.
The expansion is therefore
\begin{align}
\begin{split}
\Delta R \approx  &\frac{0.2R_\mathrm{J}\nabla_\mathrm{a}}{\left(1+b\right)w}\lrfe{L_\star}{L_{\sun}}{1/4}\lrfe{M_\star}{M_{\sun}}{-1/6}\lrfe{\tau_{\mathrm{orbit}}}{10\si{d}}{-4/3}\lrfe{M}{M_\mathrm{J}}{-1}\\
	&\times\lrfe{R}{R_\mathrm{J}}{2}\left(\frac{L_{\mathrm{i,f}}^2}{L_{\mathrm{i,i}} L_3} \right)^{-\frac{\nabla_\mathrm{a}}{w}} \ln \frac{L_{\mathrm{i,f}}}{L_\mathrm{i,i}}\ln \frac{L_2}{L_1}.
\label{eq:expansion1}
\end{split}
\end{align}
A factor of a few from the luminosity term is therefore sufficient to substantially inflate the planet at short orbital periods.

\section{G-Modes}

The existence of internal radiative zones raises the possibility that g-modes may contribute to tidal heating.
This is particularly interesting because, if g-mode dissipation is the dominant form of tidal heating $\epsilon$ is actually a function of the thermal structure of the planet.
That is because g-modes predominantly resonate in radiative zones.
What this amounts to is a form of feedback between the thermal and mechanical structures of the planet.

\subsection{Dynamical Tide}

In principle there are two sources of dynamical tides, namely gravitational and thermal.
We expect that thermal tides do not couple to the g-modes considered here.
There are two reasons for this.
First, the thermal tide is significant only in the upper layers of the atmosphere where insolation is significant.
In particular, the tide damps as $e^{-\kappa \Sigma}$\ \citep{2010ApJ...714....1A}.
The internal radiative zone begins at a comparable column density to the unperturbed radiative-convective transition.
Equation\ \eqref{eq:sigt} gives $\kappa_0 \Sigma_\mathrm{t} \approx 5\times 10^2$, so the damping is on the order of $\exp(-5\times 10^2)$, which suffices to make this effect negligible.
Secondly, the thermal tide relies on timescale for redistributing heat being large relative to the orbital time.
We have shown that the temperature distribution becomes spherical very near the photosphere and well above the convection zone, even for a tidally locked planet.
This means that it will not reach even the upper convection zone.
As a result we restrict our analysis to gravitational tides.

Due to their frequencies being small relative to the acoustic frequency, g-modes are unlikely to substantially compress material in the planet.
As a result we must treat them in the incompressible limit.
This may be done by separating the perturbing tidal potential into a hydrostatic equilibrium tide and a dynamical tide\ \citep{1975A&A....41..329Z}.
The associated radial displacements $\xi^\mathrm{eq}$ and $\xi^{\mathrm{dyn}}$ obey the relations\ \citep{1998ApJ...507..938G}
\begin{equation}
	\xi^\mathrm{eq} = -\frac{\delta \Phi}{d\Phi/dr}
	\label{eq:dphi}
\end{equation}
and
\begin{align}
	\label{eq:fulltidemode00}
	\frac{\partial^2}{\partial r^2} (r^2 \xi^\mathrm{dyn})& + \frac{\partial}{\partial r}\left(\frac{d\ln \rho}{dr} r^2 \xi^\mathrm{dyn}\right) + l(l+1)\left(\frac{N^2}{\omega^2}-1\right) \xi^{\mathrm{dyn}}\\
	& = l(l+1)\xi^{\mathrm{eq}} - \frac{\partial^2}{\partial r^2}\left(r^2\xi^{\mathrm{eq}}\right),\nonumber
\end{align}
where $l$ is the latitudinal quantum number, $\omega$ is the frequency, $\Phi$ is the unperturbed planetary gravitational potential, $\delta \Phi$ is the perturbing tidal potential due to the star and $N$ is the Brunt-V\"ais\"al\"a frequency, with $N^2$ positive in the radiative zone and negative in the surrounding convective regions.
To analyse this equation we first solve the homogeneous version, with the right hand side set to zero, and then compute the overlap between the resulting modes and the forcing term given by the right-hand side.

\subsection{Mode Profile}

The homogeneous part of equation\ \eqref{eq:fulltidemode00} is
\begin{equation}
	\frac{\partial^2}{\partial r^2} (r^2 \xi^\mathrm{dyn}) + \frac{\partial}{\partial r}\left(\frac{d\ln \rho}{dr} r^2 \xi^\mathrm{dyn}\right) + l(l+1)\left(\frac{N^2}{\omega^2}-1\right) \xi^{\mathrm{dyn}} = 0.
\end{equation}
Defining
\begin{equation}
	\xi \equiv r^2 \xi^{\mathrm{dyn}}	
\end{equation}
we find
\begin{equation}
	\frac{\partial^2 \xi}{\partial r^2} + \frac{\partial}{\partial r}\left(\frac{d\ln \rho}{dr} \xi\right) + \frac{l(l+1)}{r^2}\left(\frac{N^2}{\omega^2}-1\right) \xi = 0.
	\label{eq:diffeq}
\end{equation}
We now wish to perform a change of variables which will eliminate the first order derivative of $\xi$.
To do this, we note that
\begin{equation}
\frac{\partial^2}{\partial r^2} = \left(\frac{\partial y}{\partial r}\right)^2 \frac{\partial^2}{\partial y^2} + \frac{\partial^2 y}{\partial r^2}\frac{\partial}{\partial y}.
\end{equation}
This may be written as
\begin{equation}
\frac{\partial^2}{\partial r^2} = \left(\frac{\partial y}{\partial r}\right)^2 \frac{\partial^2}{\partial y^2} + \frac{\partial r}{\partial y} \frac{\partial^2 y}{\partial r^2}\frac{\partial}{\partial r}.
\end{equation}
Using this, we pick
\begin{equation}
	y = \int \rho^{-1} dr,
\end{equation}
which gives
\begin{equation}
\frac{\partial^2}{\partial r^2} = \rho^{-2} \frac{\partial^2}{\partial y^2} - \frac{d \ln \rho}{d r}\frac{\partial}{\partial r}.
\end{equation}
With this substitution, \eqr{eq:diffeq} becomes
\begin{equation}
	\rho^{-2}\frac{\partial^2 \xi}{\partial y^2} + \frac{d^2\ln \rho}{dr^2} \xi + \frac{l(l+1)}{r^2}\left(\frac{N^2}{\omega^2}-1\right) \xi = 0.
	\label{eq:diffeq1}
\end{equation}
Qualitatively we expect $N$ to peak near the centre of the radiative zone and fall to zero at the edges.
To fit this, we pick a quadratic form in our new coordinate $y$, such that
\begin{equation}
N^2 = N_0^2 \left(1-\left(\frac{y-y_0}{\delta y}\right)^2\right),
\end{equation}
where $r_0$ is the radial coordinate of the centre of the radiative zone and $2\delta\! y$ is the width of the zone in $y$.
Defining
\begin{align}
\Omega &\equiv \frac{N_0}{\omega}\\
\intertext{and}
x &\equiv \frac{y-y_0}{\delta y},
\end{align}
the differential equation \eqr{eq:diffeq1} becomes
\begin{equation}
	\frac{1}{\rho^2 \delta y^2}\frac{\partial^2 \xi}{\partial x^2} + \frac{d^2\ln \rho}{dr^2} \xi + \frac{l(l+1)}{r^2}\left(\Omega^2 - x^2 \Omega^2 - 1\right) \xi = 0.
	\label{eq:diffeq2}
\end{equation}
We now define
\begin{equation}
	q \equiv 1 - \frac{d^2\ln \rho}{dr^2} \frac{r^2}{l(l+1)},
\end{equation}
such that
\begin{equation}
	\frac{\rho^2}{\delta y^2}\frac{\partial^2 \xi}{\partial x^2} + \frac{l(l+1)}{r^2}\left(\Omega^2 - x^2 \Omega^2 - q\right) \xi = 0.
	\label{eq:diffeq3}
\end{equation}
Note that $q$ is positive and large because
\begin{equation}
-r^2 \frac{d^2\ln \rho}{dr^2}	 \approx \frac{r^2}{h^2} \gg 1.
\end{equation}
This follows because $\rho$ has characteristic scale $h$ and because $h \ll r$ except near the core of the planet.

It is now worth noting that the physical width of the zone in $r$ is 
\begin{equation}
	l_r \approx \rho\delta y.	
\end{equation}
This holds because for a thin zone, $\rho$ does not change too much across it.
In thick zones there would be deviations from this which we neglect.
For convenience we now define
\begin{align}
	\beta &\equiv l(l+1)\left(\frac{l_r}{r}\right)^2.
\end{align}
With this, \eqr{eq:diffeq3} becomes
\begin{equation}
	\frac{\partial^2 \xi}{\partial x^2} + \beta\left(\Omega^2 - x^2 \Omega^2 - q\right) \xi = 0.
	\label{eq:diffeq4}
\end{equation}
This may also be written as
\begin{equation}
\left(\Omega^2 - q\right)\xi = \left(\Omega^2 x^2 - \beta^{-1} \frac{\partial^2}{\partial x^2}\right)\xi
\label{eq:diffeq5}
\end{equation}
which is the same as the equation for a quantum harmonic oscillator with energy $\Omega^2-q$, mass $\hbar^2 \beta/2$, and zero-point energy $\Omega/\sqrt{\beta}$.
The eigenvalues are therefore quantised in the form
\begin{equation}
\Omega^2-q = \frac{2\Omega}{\sqrt{\beta}}\left(\frac{1}{2}+n\right),n=0,1,2,\dots .
\end{equation}
The sign of $\Omega$ does not enter into \eqr{eq:diffeq1} so we may take whichever branch of the solutions to this equation that we choose.
Taking the positive we see that
\begin{equation}
\Omega_n = \frac{1+2n+ \sqrt{1+4(\beta q+n+n^2)}}{2\sqrt{\beta}},n=0,1,2,\dots.
\label{eq:OmegaN}
\end{equation}
These correspond to periods and frequencies of
\begin{align}
T_n &= 2\pi\frac{1+2n+ \sqrt{1+4(\beta q+n+n^2)}}{2 N_0 \sqrt{\beta}},n=0,1,2,\dots\\
\intertext{and}
\label{eq:omegaN}
\omega_n &= \frac{2 N_0 \sqrt{\beta}}{1+2n+ \sqrt{1+4(\beta q+n+n^2)}},n=0,1,2,\dots.
\end{align}
The radial profiles of the solutions are the product of an exponential with an Hermite polynomial.
For $l=2$, the dominant tidal mode, these are
\begin{align}
\label{eq:eigenfunctions}
\psi_{n,m}(r,\theta,\phi) &\approx \frac{\sqrt{2\Omega_n\sqrt{\beta}}}{\sqrt{2^n n! r^3 \sqrt{\pi}}}e^{-2\Omega_n\sqrt{\beta} x^2 /2}\\
	&\times H_n\left(x\sqrt{2\Omega_n\sqrt{\beta}}\right) Y_{2m}(\theta,\phi),\nonumber
\end{align}
where $Y_{lm}$ are the spherical harmonics, $n \in \left\{0,1,2,...\right\}$ and $m\in \left\{-2,1,0,1,2\right\}$.
The modes are normalised so that
\begin{equation}
\int_{\mathrm{all\ space}} d^3\boldsymbol{r} |\psi_{n,m}|^2 = 1,
\label{eq:normalisation}
\end{equation}
and we take $\rho$ and $r$ as constants throughout the radiative zone to compute this normalisation.
This is consistent with approximations we make elsewhere.

\subsection{Overlap Integral}

In order to compute the tidal forcing $F_{n,m}(\omega)$, we must say something about the origin of the tidal potential.
There are two potential sources, rotational asynchronisation and orbital eccentricity.
In the former, the tidal forcing occurs at a frequency $\omega_\mathrm{rotation} - \Omega_\mathrm{orbit}$, while in the latter it occurs at a frequency of $\Omega_\mathrm{orbit}$.
In both cases, working in the frame corotating with the planet's orbit,
\begin{equation}
	\delta \Phi \propto \frac{G M_\star r^2}{a_\mathrm{orbit}^3}	,
\end{equation}
and further involves a sum of $l=2$ spherical harmonics.
Beyond this the two cases differ significantly because the eccentricity case has $\delta \Phi \propto e$ while the asynchronous case has no such factor.
To capture both cases, we write
\begin{equation}
	\delta \Phi = \Pi \frac{G M_\star r^2}{a_\mathrm{orbit}^3} \sum_{m'} Y_{2m'}(\theta,\phi) k_{m'} \cos(\omega t - \phi_{m'}),
\end{equation}
where $\phi_{m'}$ are phase factors,  the factors $k_{m'}$ capture the magnitudes of the various harmonics and sum in quadrature to unity and $\Pi$ is a dimensionless factor of order unity in the asynchronous case and of order $e$ in the eccentric case.
From this form and \eqr{eq:dphi} we may write the equilibrium tide as
\begin{equation}
	\xi^{\mathrm{eq}} = 	\Pi \frac{M_\star r^4}{m a_\mathrm{orbit}^3} \sum_{m'} Y_{2m'}(\theta,\phi) k_{m'} \cos(\omega t - \phi_{m'}).
\end{equation}
The driving term associated with this equilibrium tide is the right-hand side of \eqr{eq:fulltidemode00}, given by
\begin{align}
	d(\xi) &= l(l+1) \xi^{\mathrm{eq}} - \frac{\partial^2}{\partial r^2}(r^2 \xi^\mathrm{eq})= -24 \xi^{\mathrm{eq}}.
\end{align}
In computing the overlap of this with the eigenmodes of the homogeneous equation, we may treat factors of $r$ as constant, because the radiative zone ought to be thin on the scale of the planetary radius.
As a result, the projection is
\begin{align}
	\langle \psi_{2,m'} | d \rangle &= -24 \int\psi_{2,m'}(\boldsymbol{r}) \xi^{\mathrm{eq}}(\boldsymbol{r}) d^3 \boldsymbol{r} \\
								& \approx -24 \Pi \frac{M_\star r^6}{m a_\mathrm{orbit}^3} k_{m'} \int_{-l_r}^{l_r} \psi_{2,m'}(r')dr'\\
								& = -24 \Pi \frac{M_\star r^6 l_r}{m a_\mathrm{orbit}^3} k_{m'} \int_{-\infty}^{\infty} \psi_{2,m'}(x)dx\\						
								& = -24 \Pi \frac{M_\star r^{9/2} l_r s_n}{m a_\mathrm{orbit}^3 \sqrt{2^n n! \sqrt{\pi}}} k_{m'} \int_{-\infty}^{\infty} e^{-s_n^2 x^2/2} H_n(s_n x)dx\\		
								& = -24 \Pi \frac{M_\star r^{9/2} l_r}{m a_\mathrm{orbit}^3 \sqrt{2^n n! \sqrt{\pi}}} k_{m'} \int_{-\infty}^{\infty} e^{- w^2/2} H_n(w)dw,						
\end{align}
where we have centred the integral on $r_0$, the radial coordinate corresponding to $y_0$, and defined
\begin{equation}
s_n \equiv \sqrt{2\Omega_n \sqrt{\beta}}.
\end{equation}
We have also extended the integration bounds to infinity to make the computation easier because the exponential suppression in $x$ makes the precise bounds irrelevant.
Note that changing variables from $r'$ to $x$ is formally quite complicated, though in the approximation where $\rho$ changes little over the course of the zone it just produces a prefactor of $l_r$.

The integral may now be evaluated with the generating function of the Hermite polynomials,
\begin{equation}
e^{2wt-t^2} = \sum_{n=0}^\infty H_n(w) \frac{t^n}{n!},
\end{equation}
so
\begin{align}
\int_{-\infty}^\infty  e^{-w^2 /2} H_n\left(w\right) dw&= \frac{d^n}{d t^n} \int_{-\infty}^\infty \left.e^{2wt-t^2-w^2/2}dw\right|_{t=0}\\
&= \sqrt{2\pi}\frac{d^n}{dt^n}\left.\left(e^{t^2}\right)\right|_{t=0}\\
&= \frac{n!\sqrt{2\pi}}{2\Gamma\left(1+\frac{n}{2}\right)}\left(1+(-1)^n\right).
\end{align}
As a result,
\begin{equation}
	\langle \psi_{2,m'} | d \rangle \approx -24 \Pi \frac{M_\star r^{9/2} l_r \pi^{1/4}}{m a_\mathrm{orbit}^3 \Gamma\left(1+\frac{n}{2}\right)} k_{m'}\sqrt{\frac{n!}{2^{n-1}}},
	\label{eq:integral1}
\end{equation}
for even $n$ and vanishes for odd $n$.
There is complete degeneracy in both the dissipation and oscillation over $m'$, so we may form a linear combination of spherical harmonics which precisely matches the forcing term.
This amounts to summing the right hand side of \eqr{eq:integral1} in quadrature over $m'$ and taking the square root, which gives
\begin{equation}
	\langle \psi_{2} | d \rangle \approx 24 \Pi \frac{M_\star r^{9/2} l_r \pi^{1/4}}{m a_\mathrm{orbit}^3 \Gamma\left(1+\frac{n}{2}\right)}\sqrt{\frac{n!}{2^{n-1}}}.
	\label{eq:integral3}
\end{equation}
This expression gives the amplitude of the resonance.
From this stage we take it as given that $l=2$ and drop the label on $\psi$.

\subsection{Dissipation}

The square of the displacement, which is proportional to the dissipation, has maxima at a distance of order $\pm a$ from the centre of the radiative zone so, even if the dampening were uniform, we would expect the dissipation to be greatest near the edges of the zone.
In practice, convective turbulence increases the dissipation just outside the zone and this assertion is even stronger.
To evaluate the strength of this effect we turn to various linear dissipation mechanisms.
Both radiative and viscous damping are potentially relevant.
For each of these we may calculate a quality factor $Q$, giving the number of undriven cycles required for an $e$-fold reduction in strength.
These combine as
\begin{equation}
Q = \frac{1}{\frac{1}{Q_{\mathrm{rad}}} + \frac{1}{Q_{\mathrm{turb}}}}.
\end{equation}

We begin with radiative damping at finite opacity.
The quality factor of mode $n$ is of order
\begin{equation}
Q_n \approx \omega_n \tau_n \approx \frac{3}{4\pi}\left(\frac{\omega_n \lambda_n}{c}\right)\left(\frac{P}{a T^4}\right) \left(\kappa \rho \lambda_n\right),
\end{equation}
where $c$ is the speed of light and $\tau_n$, $\omega_n$ and $\lambda_n$ are the lifetime, frequency and wavelength corresponding to mode $n$\ \citep{billpress1}.
The wavelength is given by
\begin{equation}
\lambda_n \approx \frac{2 l_r}{n+1}
\end{equation}
which just comes from the fact that mode $n$ has $n+1$ nodes over the zone width of $2l_r$.
Thus
\begin{equation}
Q_n \approx \frac{3}{4\pi}\left(\frac{P}{a T^4}\right) \frac{4 \omega_n \kappa \rho l_r^2}{c(n+1)^2}.
\end{equation}
Let
\begin{equation}
m_\mathrm{z} \equiv 2 l_r \rho,
\end{equation}
an approximate zone mass.
We find
\begin{equation}
Q_n \approx \frac{3}{4\pi}\left(\frac{P}{a T^4}\right) \frac{2 \omega_n m_\mathrm{z} \kappa  l_r}{c(n+1)^2}.
\end{equation}
From \eqr{eq:scaleheight} we know that the scale height $h$ is proportional to $T$, so
\begin{equation}
\left(\frac{h}{r}\right)^4 \left(\frac{P}{a T^4}\right) \approx \frac{P k_B^4}{m_{\mathrm{p}}^4 g^4 r^4 a}.
\end{equation}
We are interested in regions which are sufficiently shallow so that $g$ is nearly constant and so
\begin{equation}
P \approx \frac{g (M-m)}{4\pi r^2},
\end{equation}
and
\begin{align}
\left(\frac{h}{r}\right)^4 \left(\frac{P}{a T^4}\right) &\approx \lrf{k_B^4}{a m_{\mathrm{p}}^4}\lrf{M-m}{4\pi r^6 g^3}\\
	&= \lrf{k_B^4}{4\pi a G^3 m_{\mathrm{p}}^4}\frac{M-m}{m^3} \approx \num{4.6e5} M_{\mathrm{J}}^2 (M-m) m^{-3}	
\end{align}
so that
\begin{equation}
Q_n \approx \num{1.1e5} \left(1-\frac{m}{M}\right)\mathcal{M}^{-2}\lrfe{M}{m}{3}\frac{2 \omega_n m_\mathrm{z} \kappa  l_r}{c(n+1)^2}\lrfe{r}{h}{4}.
\label{eq:qRad}
\end{equation}

The radiative zone is stably stratified so we expect turbulent damping to be limited to the evanescent part of the mode which leaks into the neighbouring convective zones.
The Navier-Stokes equation with a simple viscosity term is
\begin{equation}
\frac{\partial \boldsymbol{\varv}}{\partial t} + \boldsymbol{\varv}\cdot\nabla \boldsymbol{\varv} = \boldsymbol{g} - \frac{\nabla p}{\rho} + \nu \nabla^2 \boldsymbol{\varv}.
\end{equation}
We now neglect the non-linear term because we are interested in understanding the linear growth and decay of modes and expect the absolute velocities involved to be small.
Without the nonlinear term
\begin{equation}
\frac{\partial \boldsymbol{\varv}}{\partial t} = \boldsymbol{g} - \frac{\nabla p}{\rho} + \nu \nabla^2 \boldsymbol{\varv}.
\end{equation}
The balance between gravity and the pressure gradient is what gives us g-modes, so we may write the balance as
\begin{equation}
\frac{\partial \boldsymbol{\varv}}{\partial t} = \omega_n \boldsymbol{\varv} + \nu \nabla^2 \boldsymbol{\varv}.
\end{equation}
Now let
\begin{equation}
\boldsymbol{\varv}' \equiv e^{i\omega_n t} \boldsymbol{\varv}
\end{equation}
so that
\begin{equation}
\frac{\partial \boldsymbol{\varv}'}{\partial t} = \nu \nabla^2 \boldsymbol{\varv}'.
\end{equation}
The kinetic energy density
\begin{equation}
K = \frac{1}{2} \boldsymbol{\varv}^* \cdot \boldsymbol{\varv} = \frac{1}{2} \boldsymbol{\varv}'^* \cdot \boldsymbol{\varv}',
\end{equation}
where $\boldsymbol{\varv}^*$ is the complex conjugate of $\boldsymbol{\varv}$, and evolves as
\begin{equation}
\partial_\mathrm{t} K = \frac{1}{2} \nu \boldsymbol{\varv}'^* \cdot \nabla^2 \boldsymbol{\varv}' +  \frac{1}{2} \nu \boldsymbol{\varv}' \cdot \nabla^{\dagger 2} \boldsymbol{\varv}'^*,
\end{equation}
where $\nabla^\dagger$ is the adjunct gradient operator.
When the spatial derivatives are greatest in the radial direction the Laplacian just produces a factor of $(2\pi/\lambda_n)^2$ so
\begin{equation}
\partial_\mathrm{t} K =\lrfe{2\pi}{\lambda_n}{2}\nu K.
\end{equation}
Integrating this equation over the whole planet we find
\begin{align}
\frac{d}{dt} \int_0^M K dm &= \int_0^M \frac{\partial K}{\partial t} dm\\
		&= \int_0^M \lrfe{2\pi}{\lambda_n}{2}\nu K dm\\
		&= \lrfe{2\pi}{\lambda_n}{2}\int_0^M \nu K dm.
\end{align}
The viscosity $\nu$ is only significant in the convection zones on either side of the radiative zone, where turbulent viscosity dominates.
The kinetic energy density $K$ is only significant inside the radiative zone and within a few wavelengths of the zone edges on either side.
The damping integral is dominated by the region in which neither is small, so 
\begin{equation}
\frac{d}{dt} \int_0^M K dm \approx \lrfe{2\pi}{\lambda_n}{2} \frac{\lambda_n}{l_r + \lambda_n}\nu \int_\mathrm{Radiative Zone}  K dm,
\end{equation}
where $\nu$ is evaluated in the convecting regions and $2(l_r + \lambda_n)$ is roughly the radial extent of the region where the kinetic energy density is significant. 
Similarly
\begin{equation}
\int_\mathrm{Radiative Zone} K dm \approx m_\mathrm{z} K,
\end{equation}
where $K$ on the right hand side is the average kinetic energy density in the zone.
Then
\begin{equation}
\frac{d}{dt} \left(m_\mathrm{z} K\right) \approx \lrfe{2\pi}{\lambda_n}{2} \frac{\lambda_n}{l_r}\nu m_\mathrm{z} K.
\end{equation}
Because $m_\mathrm{z}$ is constant we find that
\begin{equation}
\frac{d \ln K}{dt} \approx \lrfe{2\pi}{\lambda_n}{2} \frac{\lambda_n}{l_r}\nu
\end{equation}
and the damping timescale is
\begin{equation}
\tau_{\mathrm{turb}} = \frac{\lambda_n l_r}{4\pi^2 \nu}
\end{equation}
with related quality factor
\begin{equation}
Q_n \approx \omega_n \tau_{\mathrm{turb}} = \frac{\omega_n \lambda_n l_r}{4\pi^2 \nu}.
\label{eq:qConvTau}
\end{equation}

The turbulent diffusivity is of order $v_c h$ when the convective turnover is on a timescale shorter than the forcing frequency $\omega$.
It is $\omega$ rather than $\omega_n$ that matters here because the oscillation physically takes place at the driving frequency, not the mode period.
The relevant turbulent frequency for motion over length-scale $l_\mathrm{t}$ is
\begin{equation}
\omega_{\mathrm{turb}}(l_\mathrm{t}) = \frac{v_c(l_\mathrm{t})}{l_\mathrm{t}}.
\end{equation}
Taking $v_c$ to be given by a Kolmogorov spectrum, we find
\begin{equation}
\omega_{\mathrm{turb}}(l_\mathrm{t}) = \frac{v_c(h)}{h}\lrfe{l_\mathrm{t}}{h}{-2/3}.
\end{equation}
It follows that the relevant diffusive motions are on a scale
\begin{equation}
\frac{l_\mathrm{t}}{h} = \min\left[1,\lrfe{v_c(h)}{h \omega}{3/2}\right]
\end{equation}
and corresponding diffusivity is
\begin{equation}
\nu = v_c h\min\left[1,\lrfe{v_c(h)}{h \omega}{2}\right].
\end{equation}
This is just the result of \citet{1977ApJ...211..934G} and yields a quality factor
\begin{equation}
Q_n \approx \frac{\omega_n \lambda_n l_r}{4\pi^2 v_c h}\max\left[1,\lrfe{v_c(h)}{h \omega}{-2}\right].
\label{eq:qConv}
\end{equation}
High-frequency driving leads to a high quality factor and $Q_n$ scales as $\omega_n^2$ because $\lambda_n \propto \omega_n$.
This is a weaker scaling than the radiative $Q$, which goes as $\omega_n^3 \propto (n+1)^{-3}$.
Thus at large $n$ the convective mechanism dominates.

We are often interested in the lowest $n$ because this mode is the least suppressed by overlap factors.
The convective flux of interest is generally $F_\mathrm{i}$, which \eqr{eq:fluxrat3} shows is on the order of $10^{-5}F_\mathrm{e}$.
The external flux is typically about $10^{-2} F_{\sun}$, so the relevant convective flux is on the order of $\SI{5e3}{erg.cm^{-2}.s^{-1}}$.
For a density of $\SI{e-1}{g.cm^{-3}}$ the convection speed is then
\begin{equation}
v_\mathrm{c} \approx \lrfe{F_\mathrm{i}}{\rho}{1/3} \approx \SI{30}{cm.s^{-1}}.
\end{equation}
So for a scale height of $\SI{e9}{cm}$, $v_\mathrm{c}/h \approx \SI{3e-8}{Hz}$.
We show later that we are interested in frequencies on the order of $\SI{e-6}{Hz}$.
This means that the factor $[h\omega/v_\mathrm{c}(h)]^2$  accounting for the eddy time is of order $10^5$, so the convective quality factor is $Q_1 \approx 300$.
By comparison, the fiducial radiative quality factor with the same assumptions is $Q_1 \approx 1$.
Thus we expect radiative damping to dominate by a reasonable margin unless the fluxes involved are many orders of magnitude larger or the frequencies are several orders of magnitude smaller.

\subsection{Boundaries}

We have shown that, when g-modes dominate the dissipation, $\varepsilon$ is significant primarily near the edges of the radiative zone.
It is also straightforward to show that the dissipation is symmetric because the squared mode profiles are even.
So $\varepsilon$ is an even function, the integral of which is dominated by the regions just outside the zone boundaries.
Suppose that the total luminosity produced by tides is $L_\mathrm{t}$ and that a fraction $f$ of this is produced inside the radiative zone.
In steady-state
\begin{equation}
L_2 - L_3 = f L_\mathrm{t}
\end{equation}
and
\begin{equation}
L_1 - L_2 = \frac{1}{2}(1-f)L_\mathrm{t}.
\end{equation}
Using \eqr{eq:L3_1} we find
\begin{equation}
L_3 = L_\mathrm{i} \frac{\nabla_\mathrm{a}}{\nabla_\mathrm{r}(P_\mathrm{inject})} + \frac{1}{2}(1-f) L_\mathrm{t}.
\end{equation}
If the heating is large relative to $L_\mathrm{i}$ and $f$ is small then we expect
\begin{equation}
L_3 \approx \frac{1}{2}(1-f) L_\mathrm{t}
\label{eq:L3}
\end{equation}
so
\begin{align}
\label{eq:L1L2rat}
\frac{L_1}{L_2} &\approx 2\\
\intertext{and}
\label{eq:L2L3rat}
\frac{L_2}{L_3} &\approx 1.
\end{align}
Using equations \eqref{eq:trat1} through \eqref{eq:prat2} we find
\begin{align}
\ln \frac{T_1}{T_2} &= \frac{\nabla_\mathrm{a}}{w}\ln 2,\\
\ln \frac{T_2}{T_3} &= \frac{\nabla_\mathrm{a}}{w}\ln \frac{L_\mathrm{t}}{L_\mathrm{i,i}},\\
\ln \frac{P_1}{P_2} &= \frac{1}{w}\ln 2\\
\intertext{and}
\ln \frac{P_2}{P_3} &= \left(\frac{1}{w} + \frac{1}{1+b}\right)\ln \frac{L_\mathrm{t}}{L_\mathrm{i,i}}.
\end{align}
With equations \eqref{eq:L3} and \eqref{eq:L1L2rat} \eqr{eq:expansion1} yields
\begin{align}
\begin{split}
\frac{\Delta R}{R_\mathrm{J}} \approx  &-0.1\frac{\nabla_\mathrm{a}}{\left(1+b\right)\left(w\right)}\lrfe{L_\star}{L_{\sun}}{1/4}\lrfe{M_\star}{M_{\sun}}{-1/6}\lrfe{\tau_\mathrm{orbit}}{10\si{d}}{-4/3}\\
	&\times\lrfe{M}{M_\mathrm{J}}{-1}\lrfe{R}{R_\mathrm{J}}{2}\left(\frac{2 L_{\mathrm{i,f}}}{L_{\mathrm{i,i}}} \right)^{\frac{-\nabla_\mathrm{a}}{w}} \ln \frac{L_{\mathrm{i,f}}}{L_\mathrm{i,i}}.
\label{eq:expansionBoundary0}
\end{split}
\end{align}

Recall from \eqr{eq:omegaN} that the resonant frequencies are
\begin{align}
\omega_n &= \frac{2 N_0 \sqrt{\beta}}{1+2n+\sqrt{1+4(\beta q+n+n^2)}}\\
		&= \frac{2N_0 \sqrt{\beta}}{\left(1+2n\right)\left(1+\sqrt{1+\frac{4\beta q}{(1+2n)^2}}\right)}\\
		&= \frac{2N_0 \sqrt{\beta}}{\left(1+2n\right)\left(1+\sqrt{1+\frac{4(l_r/h)^2}{(1+2n)^2}}\right)}.
\end{align}
The right hand side has two characteristic regimes, one in which $l_r/h$ is large and one in which it is small or of order unity.
In the former case, $\omega_n \propto h$, while in the latter $\omega_n \propto l_r$.
Both $h$ and $l_r$ increase with $P_3/P_2$, so the resonant frequency goes up as the zone width increases.
This means that the resonance shifts up in frequency as the luminosity increases.
That is,
\begin{equation}
\frac{d\omega_0}{d L_\mathrm{t}} > 0.
\end{equation}
So an increase in luminosity tends to tune the system towards resonance if it is being driven above resonance and pushes it away from resonance otherwise.
The net result is that there is thermomechanical feedback which tends to bias systems towards resonance, particularly when their resonant frequency is below the driving frequency.

\subsection{Power Production}

Each mode may be treated as a separate damped and forced harmonic oscillator.
Let $\xi_{n}$ be the amplitude for mode $n$.
Then
\begin{equation}
\omega_n^2 \xi_{n} + \frac{\omega_n}{Q_n} \dot{\xi}_{n} + \ddot{\xi}_{n} = \omega^2\langle \psi | d \rangle e^{i\omega t}.
\end{equation}
We may solve this differential equation in steady-state and fix the reference phase to find
\begin{equation}
\xi_n = \frac{\omega^2\langle \psi | d \rangle e^{i\omega t}}{\omega_n^2-\omega^2 + i\omega_n \omega Q_n^{-1}}.
\end{equation}
The power dissipated is
\begin{align}
\mathcal{P}_n &= \rho\Re\left(\dot{\xi}_n^{*} e^{i\omega t}\omega^2\langle \psi | d \rangle\right)\\
&= \rho\Re\left(-i \xi_n^{*}  e^{i\omega t}\right) \omega^3 \langle \psi | d \rangle\\
&= \rho\Im\left(\xi_n^{*}  e^{i\omega t}\right) \omega^3 \langle \psi | d \rangle\\
&= \rho\Im\left(\frac{1}{\omega_n^2-\omega^2 - i\omega_n \omega Q_n^{-1}}\right) \omega^5 |\langle \psi | d \rangle|^2\\
&= \frac{\omega_n \omega Q_n^{-1}}{(\omega_n^2-\omega^2)^2 + (\omega_n \omega Q_n^{-1})^2} \rho\omega^5 |\langle \psi | d \rangle|^2.
\end{align}
With \eqr{eq:integral3} this becomes
\begin{align}
\mathcal{P}_n &= \frac{\omega_n \omega^6 Q_n^{-1}}{(\omega_n^2-\omega^2)^2 + (\omega_n \omega Q_n^{-1})^2}q\\ 
&= \frac{\omega^3}{Q_n \omega_n^{-1} \omega \left(\frac{\omega_n^2}{\omega^2}-1\right)^2 + \omega^{-1} \omega_n Q_n^{-1}}q,
\label{eq:powerdissip0}
\end{align}
where
\begin{equation}
	q \equiv 576 \Pi^2 \rho\frac{M_\star^2 r^{9} l_r^2 \pi^{1/2}}{m^2 a_\mathrm{orbit}^6}\left(\frac{n!}{2^{n-1}\Gamma\left(1+\frac{n}{2}\right)^2}\right).	
\end{equation}

If the tides are driven by the rotational energy of the planet then $\omega$ is the planet's rotation frequency.
In many cases however the tides are driven by either orbital eccentricity, in which case the forcing frequency is just the orbital frequency\ \citep{2010ApJ...714....1A}.
In this more generally interesting case, the driving frequency is
\begin{equation}
\omega = \Omega_\mathrm{orbit} = \frac{2\pi}{\tau_\mathrm{orbit}} \approx \SI{7e-6}{Hz}\lrfe{\tau_\mathrm{orbit}}{\SI{10}{d}}{-1}.
\end{equation}
To compare, the highest resonant frequency occurs when $n=0$ and is
\begin{equation}
\omega_0 = \frac{2 N_0 \sqrt{\beta}}{1+\sqrt{1+4\beta q}},
\end{equation}
\eqr{eq:omegaN}.
Because $\beta q \approx 1$,
\begin{equation}
\omega_0 \approx \frac{N_0 \sqrt{\beta}}{1+\sqrt{5}} \approx \frac{N_0 l_r}{R},
\end{equation}
for $l=2$.
Now $N_0$ is the peak Brunt-V\"ais\"al\"a frequency in the radiative zone where the temperature gradient is substantially subadiabatic, so
\begin{equation}
N_0^2 \approx \nabla_\mathrm{a} \frac{g}{h}.
\end{equation}
This gives
\begin{equation}
\omega_0 \approx \frac{l_r\sqrt{\nabla_\mathrm{a}}}{R}\sqrt{\frac{g}{h}}.
\end{equation}
Now $l_r \approx h \ln (P_1/P_\mathrm{t,i})$ so
\begin{align}
\omega_0 &\approx \sqrt{\nabla_\mathrm{a}} \ln \frac{P_1}{P_\mathrm{t,i}}\sqrt{\frac{g h}{R^2}}\\
 	&=\frac{1}{\sqrt{\gamma}}\lrf{c_\mathrm{s}}{R} \ln \frac{P_1}{P_\mathrm{t,i}}\\
	&\approx \SI{2e-5}{Hz}\lrfe{T}{\SI{1e3}{K}}{1/2}\lrfe{R}{R_\mathrm{J}}{-1} \ln \frac{P_1}{P_\mathrm{t,i}}.
\end{align}
This is quite close to the orbital frequency so we expect that resonances are not uncommon.
Note also that it is somewhat greater than the orbital frequency, so there will generally be modes with frequencies lower than the orbital frequency which are pulled upward towards it by thermomechanical feedback.

The precise shape and spacing of the resonances depends on our ansatz for the Brunt-V\"ais\"al\"a frequency in the radiative zone and so it is not useful to make predictions which depend on our chosen form.
If we instead average over resonances, we note that the forcing integral falls exponentially in $n$ while the resonances fall off as a power law in $n$ so that the $n=0$ resonance always dominates on average.
Individual systems far from the $n=0$ resonance may exhibit significant dissipation by virtue of sitting directly on a higher resonance but we expect this to be rare.
So we assume that $\omega$ and $\omega_n$ are of the same order and write the net power
\begin{align}
\mathcal{P} &= \sum_i \mathcal{P}_i\\
			&\approx \mathcal{P}_0\\
			&\approx 2\times 10^3 Q_0^{-1}\Pi^2 \rho\frac{\Omega_\mathrm{orbit}^3 M_\star^2 r^{9} l_r^2}{m^2 a_\mathrm{orbit}^6}\left(\frac{\omega_0}{\Omega_\mathrm{orbit}}\right).	
\end{align}
This may be converted to a flux as
\begin{equation}
F_\mathrm{t} = \frac{\mathcal{P}}{4\pi r^2} \approx 2\times 10^2 Q_0^{-1}\Pi^2 \rho\frac{\Omega_\mathrm{orbit}^3 M_\star^2 r^{7} l_r^2}{m^2 a_\mathrm{orbit}^6}\left(\frac{\omega_0}{\Omega_\mathrm{orbit}}\right).
\label{eq:tidalFlux}
\end{equation}

\section{Equilibrium Radius}

To first order, suppose that $l_r \approx h$.
The tidal flux is then given by
\begin{align}
F_\mathrm{t} \approx 2\times 10^2 Q_0^{-1}\Pi^2 \rho\frac{\Omega_\mathrm{orbit}^3 M_\star^2 r^{7} h^2}{m^2 a_\mathrm{orbit}^6}\left(\frac{\omega_0}{\Omega_\mathrm{orbit}}\right).
\end{align}
Inserting \eqr{eq:qRad} yields
\begin{align}
F_\mathrm{t} \approx 2\times 10^{-3} \Pi^2 \rho\frac{\Omega_\mathrm{orbit}^3 m \mathcal{M}^{2} r^3 h^5 M_\star^2 c}{2 \omega_0 m_\mathrm{z}\kappa M^3 a_\mathrm{orbit}^6}\left(\frac{\omega_0}{\Omega_\mathrm{orbit}}\right)\left(1-\frac{m}{M}\right)^{-1}.	
\end{align}
If the radiative zone is shallow but dominates the mass above its base then $m \approx M$.
In addition, 
\begin{equation}
	M - m \approx 4\pi r^2 \rho l_r \approx 4\pi r^2 \rho h
\end{equation}
so
\begin{align}
F_\mathrm{t} \approx &\ 2\times 10^{-3} \Pi^2 \rho\frac{r^3 h^5 \Omega_\mathrm{orbit}^2 M \pi^{1/2}c}{8\pi r^2 \rho^2 h^2 \kappa a_\mathrm{orbit}^6} \lrfe{M_\star}{M_\mathrm{J}}{2}\\
\approx &\ 2\times 10^{3} \Pi^2 \rho\frac{r h^5 \Omega_\mathrm{orbit}^2 M c}{8\pi (\rho h)^2 \kappa a_\mathrm{orbit}^6} \lrfe{M_\star}{M_{\sun}}{2}.
\end{align}
Noting that
\begin{equation}
	\Sigma_\mathrm{t} \approx h \rho	
\end{equation}
we find
\begin{align}
F_\mathrm{t} \approx &\ 8\times 10^{1} \Pi^2\frac{r h^4 \Omega_\mathrm{orbit}^2 M c}{\Sigma_\mathrm{t}  \kappa a_\mathrm{orbit}^6} \lrfe{M_\star}{M_{\sun}}{2}.
\end{align}
Recalling \eqr{eq:sigt} and neglecting the logarithmic correction to $\Sigma_\mathrm{t}$ owing to the motion of the zone boundary we write
\begin{align}
F_\mathrm{t} \approx &\num{5e-5}F_\odot\Pi^2\lrfe{\tau_\mathrm{orbit}}{\SI{10}{d}}{-\frac{31}{6}}\lrfe{M_\star}{M_{\sun}}{\frac{5}{12}}\lrf{R_0}{R_\mathrm{J}}\nonumber\\
				&\times\mathcal{R}^{-2}\mathcal{M}^{8/3}\lrfe{L_\star}{L_{\sun}}{-\frac{5}{8}}\psi^{\frac{5}{2}}\lrfe{\kappa}{\kappa_0}{-1}\lrfe{h}{r}{4}.
				\label{eq:fluxfinal}
\end{align}
With the fiducial values and $h \approx 0.2 r$ this flux produces an expansion at a rate
\begin{align}
	\frac{dR}{dt} &\approx \frac{F_\mathrm{t}}{\rho c_\mathrm{p} T} \approx \frac{F}{P_\mathrm{t}} \approx \SI{3e-5}{cm.s^{-1}} \Pi^2,
\end{align}
which is sufficient to produce expansion of order $R_\mathrm{J}$ over million-year timescales.

As discussed in section\ \ref{sec:expansion} the expansion eventually increases the escaping flux to match the generated flux, so the expansion does not continue forever.
The relevant dimensionless parameter for this equilibrium is the ratio $F_\mathrm{t}$ to $F_\mathrm{i,i}$, which is the unperturbed $F_\mathrm{i}$.
Recall from \eqr{eq:fluxrat3} that
\begin{align}
F_{\mathrm{i}} &\approx \num{5e-4}F_\mathrm{e} \mathcal{M}^{5/3}\lrfe{L_\star}{L_{\sun}}{-5/8}\lrfe{a_\mathrm{orbit}}{R_{\sun}}{5/4}\lrfe{\kappa}{\kappa_0}{-1}\frac{\psi^{5/2}}{\mathcal{R}^3}.
\end{align}
Inserting \eqr{eq:fefsun} yields
\begin{align}
F_{\mathrm{i}} &\approx \num{1.6e-4}F_{\sun} \mathcal{M}^{5/3}\lrfe{L_\star}{L_{\sun}}{3/8}\lrfe{a_\mathrm{orbit}}{R_{\sun}}{-3/4}\lrfe{\kappa}{\kappa_0}{-1}\frac{\psi^{5/2}}{\mathcal{R}^3}\\
			&\approx \num{3e-6}F_{\sun} \mathcal{M}^{5/3}\lrfe{L_\star}{L_{\sun}}{3/8}\lrfe{\tau_\mathrm{orbit}}{\SI{10}{d}}{-1/2}\\
			&\times\lrfe{M_\star}{M_{\sun}}{-1/4}\lrfe{\kappa}{\kappa_0}{-1}\frac{\psi^{5/2}}{\mathcal{R}^3}.\nonumber
\end{align}
Thus
\begin{align}
\frac{2F_\mathrm{t}}{F_\mathrm{i,i}} \approx &\ \num{3e7}\Pi^2\mathcal{M}\mathcal{R}^{-2}\lrfe{L_{\star}}{L_{\sun}}{-1}\lrf{R_0}{R_\mathrm{J}}\\
	&\times\lrfe{\tau_\mathrm{orbit}}{\SI{10}{d}}{-14/3}\lrfe{M_\star}{M_{\sun}}{2/3}\lrfe{h}{R}{4}.\nonumber
\end{align}
Inserting \eqr{eq:scaleHeightExp} we get
\begin{align}
\frac{2F_\mathrm{t}}{F_\mathrm{i,i}} \approx &\ \num{5e4}\Pi^2\mathcal{M}^{-3}\mathcal{R}^{2}\lrfe{R_0}{R_\mathrm{J}}{5}\lrfe{\tau_\mathrm{orbit}}{\SI{10}{d}}{-10}\lrfe{L_{\mathrm{i,f}}^2}{L_{\mathrm{i,i}} L_3}{\frac{-7\nabla_\mathrm{a}}{w}}.
\end{align}
If the expansion is small then the luminosity ratios may be replaced by flux ratios.
If tidal heating is significant, the perturbed flux escaping from the interior of the planet $F_\mathrm{i,f} \approx F_\mathrm{t}$, so
\begin{align}
\frac{2F_\mathrm{t}}{F_\mathrm{i,i}} \approx &\ \num{5e4}\Pi^2\mathcal{M}^{-3}\mathcal{R}^{2}\lrfe{R_0}{R_\mathrm{J}}{5}\lrfe{\tau_\mathrm{orbit}}{\SI{10}{d}}{-10}\lrfe{F_{\mathrm{i,f}}^2}{F_{\mathrm{i,i}} F_3}{\frac{-7\nabla_\mathrm{a}}{w}}.
\end{align}
Using equations \eqref{eq:L1L2rat} and \eqref{eq:L2L3rat} we find
\begin{align}
\frac{2F_\mathrm{t}}{F_\mathrm{i,i}} \approx &\ \num{5e4}\Pi^2\mathcal{M}^{-3}\mathcal{R}^{2}\lrfe{R_0}{R_\mathrm{J}}{5}\lrfe{\tau_\mathrm{orbit}}{\SI{10}{d}}{-10}\lrfe{2 F_{\mathrm{i,t}}}{F_{\mathrm{i,i}}}{\frac{-7\nabla_\mathrm{a}}{w}}.
\label{eq:fluxratTides}
\end{align}
Note that $R$ and $\mathcal{R}$ must be the equilibrium radii here, not pre-expansion radii.
Note that the absolute magnitude of the opacity does not enter our final expression.
This cancellation is due to the assumption of efficient convection.

The quantity
\begin{equation}
f \equiv -{\frac{7\nabla_\mathrm{a}}{w}}
\end{equation}
is of key importance to the nature of the solution.
If $f < 1$ then the solution is stable, meaning that a system initially perturbed away from this equilibrium solution returns to it over time.
What this means physically is that an increase in the generated flux increases the temperature at the base of the radiative zone enough that the flux which escapes increases by more, leading to a negative feedback loop.
If $f > 1$ then the solution is unstable, meaning that an increase in the generated flux increases the temperature in the radiative zone by less than what is required to allow that additional flux to escape, leading to a positive feedback loop.
If a system has $f > 1$ then either the initial perturbation has the radiative zone deep enough that the runaway process keeps increasing the flux until some of our assumptions break down or the initial perturbation has the radiative zone shallow enough that the runaway process prevents it from migrating inward causing it to stay where it initially forms.
If the prefactor on the right side of \eqr{eq:fluxratTides} exceeds unity then the unstable branch is always the relevant one because the initial perturbation must yield a ratio of at least unity in order to cause a radiative zone to form.
If the prefactor is less than one then any radiative zone formed simply stays where the initial perturbation produces it.

Despite the uncertainties in the precise numbers involved, our results are fairly robust because the right-hand side of \eqr{eq:fluxratTides} scales as $\tau_\mathrm{orbit}^{-10}$, and this scaling only becomes stronger once $f$ is taken into account.
For orbital periods shorter than of about $\SI{10}{d}$, with some uncertainty, the flux generated may exceed the flux escaping from the centre of the planet.
When the tidal flux dominates the expansion is given by \eqr{eq:expansionBoundary0} and may be approximated by
\begin{align}
\frac{\Delta R}{R_\mathrm{J}} \approx  &0.1\frac{\nabla_\mathrm{a}}{\left(1+b\right)\left(w\right)}\lrfe{L_\star}{L_{\sun}}{1/4}\lrfe{M_\star}{M_{\sun}}{-1/6}\lrfe{\tau_\mathrm{orbit}}{10\si{d}}{-4/3}\nonumber\\
	&\times\lrfe{M}{M_\mathrm{J}}{-1}\lrfe{R}{R_\mathrm{J}}{2}\left(\frac{2 F_{\mathrm{i,f}}}{F_{\mathrm{i,i}}} \right)^{\frac{-\nabla_\mathrm{a}}{w}} \ln \frac{F_{\mathrm{i,f}}}{F_\mathrm{i,i}}.
\label{eq:expansionBoundary1}
\end{align}
Neglecting the logarithmic dependence, this may be combined with \eqr{eq:fluxratTides} to give
\begin{align}
\frac{\Delta R}{R_\mathrm{J}} \approx  &0.1\frac{\nabla_\mathrm{a}}{\left(1+b\right)\left(w\right)}\lrfe{L_\star}{L_{\sun}}{1/4}\lrfe{M_\star}{M_{\sun}}{-1/6}\lrfe{\tau_\mathrm{orbit}}{10\si{d}}{-\left(\frac{4}{3}+\frac{10f}{7(1-f)}\right)}\nonumber\\
	&\times\lrfe{M}{M_\mathrm{J}}{-1}\lrfe{R}{R_\mathrm{J}}{2}\left(\num{5e4}\Pi^2\mathcal{M}^{-3}\mathcal{R}^{2}\lrfe{R_0}{R_\mathrm{J}}{5}\right)^{\frac{f}{7(1-f)}}.
\label{eq:expansionBoundary2}
\end{align}
Even though the dependence on the flux ratio is small, the ratio itself can be quite large, particularly at smaller periods.
Many cases, such as $a=0,b=2$ or $a=1,b=1$, have $f > 1$ and so orbital periods of order $\SI{30}{d} \Pi^{1/5}$ suffice to cause expansion of order $R_0$.
It is difficult to say more because many of our approximations break down at this point.
If $f<1$, as can be achieved for example with $a=4,b=2$, $\Delta R/R \propto \tau_\mathrm{orbit}^{-64/3}$, orbital periods of order $\SI{20}{d}\Pi^{3/16}$ suffice to produce unit expansion.
This is consistent with most of the known cases of highly inflated Jupiter-mass planets, assuming $\Pi \approx e \approx 0.1$.
For low-mass planets, the lower surface gravity makes larger expansion more feasible.
This has been recently observed\ \citep{1606.04556}.
Stronger claims are difficult to make analytically because the dependence of the flux ratio on the specifics of the opacity are quite severe and the detailed compositions of the atmospheres of exoplanets at intermediate depths remain largely unknown.
Precision studies of this thermomechanical feedback will likely require numerical tools in all but the simplest cases.

At sufficiently low masses, large flux ratios become impossible to attain given the factor of $\mathcal{M}^{-3}$ in \eqr{eq:fluxratTides}.
At this point further expansion is impossible.
Likewise at large enough radii the central adiabat disappears so that much of this analysis becomes invalid.
We do not expect this to be a limiting factor, however.
At short periods Roche lobe overflow becomes a substantial barrier.
Substantial changes in the opacity may also occur, particularly if the Kramer regime becomes relevant, and this may invalidate much of the analysis too.
In addition at large radii the neglected factors of $R$ in converting from luminosities to fluxes become relevant and these act to limit the expansion. 

\section{Energetic Timescales}

If the tides are eccentricity-driven, it is important to consider the timescale over which the orbit circularises.
It suffices to the level of accuracy of interest to note that the energy which may be extracted from an orbit of eccentricity $e < 1$ is of order $e^2 \varv_\mathrm{orbit}^2 M$.
The circularisation timescale is therefore
\begin{equation}
	\tau_\mathrm{circ} \approx \frac{e^2 \varv_\mathrm{orbit}^2 M}{\mathcal{P}} \approx \lrfe{\tau_\mathrm{orbit}}{10\si{d}}{5/2} \SI{2e12}{yr},
\end{equation}
where we have taken $\Pi \approx e$ and used our fiducial values for all parameters other than $h/r$, which we have taken to be $0.2$.
From this it is clear that most systems of interest can be eccentricity-driven for billion-year timescales, even if they require shorter periods than the fiducial.

If the tides are driven by the planet's rotation they generally have many orders of magnitude less energy to draw from, and so are not sustained on the timescales of interest.
They may still produce bloating, but not for long enough to be easily observable.

\section{Comparison}

For comparison with other mechanisms it is useful to compute the Love number associated with these g-modes.
The Love number is defined in terms of the power and perturbing tidal potential as
\begin{equation}
	\Im[k_l^m(\omega)] = \frac{8\pi G \mathcal{P}}{(2l+1) r |\delta \Phi|^2 \omega}
\end{equation}
\citep{2014ARA&A..52..171O}.
Summing in quadrature over all $m$ and using $l=2$ and equation\ \eqref{eq:fluxfinal} we find
\begin{align}
	\Im(k) \approx& \num{3e-6}\left(\frac{\tau_{\mathrm{orbit}}}{10\mathrm{d}}\right)^{-17/6}\left(\frac{M_\star}{M_{\sun}}\right)^{-11/12}\nonumber\\
	&\times\mathcal{R}^{-3}\mathcal{M}^{8/3}\left(\frac{L_{\star}}{L_{\sun}}\right)^{-5/8}\psi^{5/2}\left(\frac{\kappa}{\kappa_0}\right)^{-1}\left(\frac{h}{0.2 r}\right)^4.
\end{align}
For comparison, inertial waves result in a frequency-averaged value of
\begin{align}
	\Im(k) &\approx 5 \frac{R^3 \Omega^2}{GM}\left(\frac{R_{\mathrm{c}}}{R}\right)^5\\
	&\approx 8\times 10^{-9}\left(\frac{M}{M_{\mathrm{J}}}\right)^{-1}\left(\frac{R}{R_{\mathrm{J}}}\right)^{3}\left(\frac{\tau_{\mathrm{orbit}}}{10\mathrm{d}}\right)^{-2}\left(\frac{R_\mathrm{c}}{0.1 R}\right)^5
\end{align}
\citep{2013MNRAS.429..613O}, where $R_\mathrm{c}$ is the core radius.
Likewise viscoelastic dissipation in a solid core potentially yields $\Im(k)\approx 10^{-6}$, scaling with the elastic properties of the core \citep{2014A&A...566L...9G, 2012A&A...541A.165R} and once more averaging over frequency.
These mechanisms are therefore of comparable order, depending on precisely which fiducial value is chosen.
The reason the g-mode mechanism inflates planets more readily despite having comparable or somewhat less power dissipation is that it heats primarily near the surface where the radius is more easily perturbed.

\section{Conclusions}

We have characterised the response of heavily insolated Jupiter-like planets to tidal heating for a wide range of tidal heating models.
A necessary condition for significant bloating of these planets is deep heating.
We find that tidal heating, either directly through tide--core interactions or indirectly through resonance-sensitive migration of radiative zones induced by g-modes, is of the right order of magnitude to induce the observed bloating if it is sufficiently deep.
We have further shown that nearly every tidal heating model results in deep heating so long as the atmosphere is sufficiently irradiated.
This explains the observation of substantially bloated hot Jupiters with a physically reasonable orbital period cutoff for such effects.
The migration of interior radiative zones provides a natural explanation for the matching of tidal frequency with orbital frequency despite the observed wide range of orbital frequencies of hot Jupiters.

This entire analysis hinges on there being a luminosity perturbation to start.
This luminosity then produces a self-sustaining interior radiative zone which dissipates substantially more heat.
The initial perturbation may come from non-linear instabilities and so may provide an indirect probe of these effects.
It may also come from planetary migration.
If a planet migrates inward and if opacity falls as a result, the incident flux can temporarily force the creation of a radiative zone while the convection zone adjusts to the reduced flux it must carry.
The g-mode hysteresis described in this paper may then prevent the zone from disappearing, even if its location shifts to better match resonance.
Inflated planets may therefore carry a record of their migration histories.

Finally, the thermomechanical feedback mechanism we propose highlights the importance of considering dynamical effects across many timescales.
Feedback is possible both from short timescales to long, as in tidal heating, and from long timescales to short, as in the dynamical tuning of g-modes.
By their very nature couplings across so many scales are difficult to track down and so there may be many more which have yet to be discovered.

\section*{Acknowledgements}

We thank Sterl Phinney for productive conversations on planetary atmospheres and heated planets.
ASJ acknowledges support from the Goldwater scholarship and the Marshall scholarship.
CAT thanks Churchill College for his fellowship.








\bibliography{refs}

\bsp	
\label{lastpage}
\end{document}